%% file: main.tex
\pdfoutput=1
\documentclass[conference]{IEEEtran}
\IEEEoverridecommandlockouts

\usepackage{cite}
\usepackage{amsmath,amssymb,amsfonts}
\usepackage{graphicx}
\usepackage{textcomp}
\usepackage{xcolor}
\usepackage{soul}
\usepackage{hyperref}
\usepackage{multirow}
\usepackage{booktabs}
\usepackage{adjustbox}
\usepackage{dirtytalk}
\usepackage[most]{tcolorbox}
\usepackage{float}
\usepackage{rotating}
\usepackage{algorithm}
\usepackage[noend]{algpseudocode}
\usepackage{ragged2e}
\usepackage{setspace}
\usepackage{array}
\usepackage{svg}
\usepackage[numbers]{natbib}

\newcommand{\thickhline}{%
    \noalign {\ifnum 0=`}\fi \hrule height 1pt
    \futurelet \reserved@a \@xhline
}

\newcolumntype{"}{@{\hskip\tabcolsep\vrule width 1.7pt\tabcolsep}}
\usepackage{caption}
\usepackage{subcaption}
	
\usepackage{color, colortbl}

\definecolor{Gray}{gray}{0.9}
\newcommand{\revision}[1]{{\textcolor{black} {#1}}}
\newcommand{\cameraready}[1]{{\textcolor{black} {#1}}}

\usepackage{tikz}
\usepackage{scalerel}
\usepackage[most]{tcolorbox}
\usepackage{setspace}
\usepackage{indentfirst}

\definecolor{main}{HTML}{5989cf}    
\definecolor{sub}{HTML}{cde4ff}     

\newtcolorbox{boxA}{
    fontupper = \bf,
    boxrule = 1.5pt,
    colframe = black 
}

\newtcolorbox{boxC}{
    colback = sub, 
    boxrule = 0pt  
}

\tcbset{
    enhanced,
    breakable,
    left=1mm,
    right=1mm,
    attach boxed title to top left={
        xshift=0.5cm,
        yshift= -3.5mm
    },
    top=3mm,
    bottom=0pt,
    colframe=blue!75!black,
    coltitle=white,
    colbacktitle=black!70!white,
    fonttitle=\footnotesize\bfseries,
    fontupper=\footnotesize\sffamily,
}

\newenvironment{mybox}[1]{%
    \begin{tcolorbox}[title={#1}]%
    \setstretch{0.95}}{
    \end{tcolorbox}
}

\makeatletter

\def\BibTeX{{\rm B\kern-.05em{\sc i\kern-.025em b}\kern-.08em
    T\kern-.1667em\lower.7ex\hbox{E}\kern-.125emX}}

\begin{document}

\title{Mutiny! How does Kubernetes fail, \\and what can we do about it?}




\author{\IEEEauthorblockN{
{Marco Barletta}\IEEEauthorrefmark{1},
{Marcello Cinque}\IEEEauthorrefmark{1},
{Catello Di Martino}\IEEEauthorrefmark{2}, 
{Zbigniew T. Kalbarczyk}\IEEEauthorrefmark{3},
and
{Ravishankar K. Iyer}\IEEEauthorrefmark{3}
}
\\
\IEEEauthorrefmark{1}{\textit{Università degli Studi di Napoli Federico II}, Naples, NA 80125, Italy}\\
\IEEEauthorrefmark{2}{\textit{Nokia Bell Labs}, Sao Paulo, SP 05069-010, Brazil}\\
\IEEEauthorrefmark{3}{\textit{University of Illinois at Urbana-Champaign}, Urbana-Champaign, IL 61801, USA}\\
}


\newcommand{\lmttfont}{\fontfamily{lmtt}\selectfont}

\newcommand{\mb}[1]{{\textcolor{green} {M: \textbf{#1}}}}
\newcommand{\mc}[1]{{\textcolor{blue} {Marcello: \textbf{#1}}}}
\newcommand{\shc}[1]{{\textcolor{red} {Shengkun: \textbf{#1}}}}
\newcommand{\hs}[1]{{\textcolor{violet} {Harshitha: \textbf{#1}}}}


\maketitle
\begin{abstract}
In this paper, we i) analyze and classify real-world failures of Kubernetes (the most popular container orchestration system), ii) develop a framework to perform a fault/error injection campaign targeting the data store preserving the cluster state, and iii) compare results of our fault/error injection experiments with real-world failures, showing that our fault/error injections can recreate many real-world failure patterns.
The paper aims to address the lack of studies on systematic analyses of Kubernetes failures to date.

Our results show that even a single fault/error (e.g., a bit-flip) in the data stored can propagate, causing cluster-wide failures (3\% of injections), service networking issues (4\%), and service under/overprovisioning (24\%). 
Errors in the fields tracking dependencies between object caused 51\% of such cluster-wide failures.
We argue that controlled fault/error injection-based testing should be employed to proactively assess Kubernetes' resiliency and guide the design of failure mitigation strategies. 
\end{abstract}


\input{1_introduction}

\input{2_background}

\input{3_ffda}

\input{4_methods}

\input{5_results}

\section{Discussion}
\label{sec:discussion}
\input{6_discussion}

\section{Related Work}
\label{sec:related}
\input{7_related}

\input{7_conclusion}




\input{main.bbl}

\end{document}

%% file: 1_introduction.tex
\section{Introduction}
\label{sec:intro}
\let\thefootnote\relax\footnotetext{
\copyright 2024 IEEE. Personal use of this material is permitted. Permission from IEEE must be
obtained for all other uses, in any current or future media, including
reprinting/republishing this material for advertising or promotional purposes, creating new
collective works, for resale or redistribution to servers or lists, or reuse of any copyrighted
component of this work in other works.}
Container orchestration systems manage container life cycles across a cluster of nodes providing automation and flexibility for application management and acting as cloud operating systems \cite{khan2017key, rodriguez2019orchestration, burns2016borg}.
They are increasingly used in mission-critical scenarios with strict non-functional requirements (e.g., response-time latency and availability) \cite{vxworks,mellado2022design,barletta2023criticality, johansson22kubernetes,nehpio,ferguson2023corekube}. Those scenarios adopt fog/edge-cloud and service-based paradigms to support fast and automated reconfiguration 
\cite{botez2021sdn,colombo2010factory,morgan2021industry,lasi2014industry,ray2019edge}.

Kubernetes (henceforth, \textit{K8s}), 
is the \textit{de facto} standard among container orchestration systems to provide automated management of services.
Several orchestration systems are compliant with K8s \cite{certifiedkube} or reuse its code.  A survey from the Cloud Native Computing Foundation found that 96\% of organizations are using (or plan to use) K8s \cite{cncf-survey}.

The widespread use of K8s-based platforms to host critical applications makes it imperative to study K8s's resiliency and failure modes. Literature has focused on K8s testing to discover bugs, but there are no available studies to assess the resiliency of K8s in a thorough and systematic way or that perform a clear classification of its failures.

K8s is designed to withstand common errors and failures through a range of resiliency strategies, including heartbeats, redundancy, failover, circuit breakers, and stateless system components.
In particular, stateless components foster easy failover: action is based on observation rather than a state machine. Upon restart, a component only needs to observe the current and the desired cluster state stored on a data store. The state dependency is thus moved away from the components to an external data store, which becomes a dependability bottleneck that preserves all the state information.

In this paper, we use both real-world failure reports and a fault/error injection campaign to classify K8s's failure modes, assess its ability to tolerate faults, and analyze how errors propagate and impact on the availability and response times of deployed services.

Multiple real-world failure reports referring to enterprise production clusters in the order of a thousand nodes \cite{kubecon5} can be found in blogs or forums\cite{k8sfailures,kubecon1,kubecon2,redditpiday,failuregke,gkefailures} on the Internet. Although human-generated reports can be vague and incomplete, they provide valuable insights to inform the design of the fault/error injection campaign.

Our fault/error injections target the data store, which preserves the current and desired states of cluster resources. 
Any corruption of the data in the data store may propagate and cause failures in every system component.
The fault/error injection campaign affects the data 
used by orchestration operations
and follow three models: bit-flips, data-type corruption, and message drops. 
Importantly, we show that our injections trigger failure patterns similar to those seen in the real world, despite the possibly different root causes.
For example, corrupting the data of a service causes containers to be spawned in an infinite loop, leading to an overload and system-wide outage. The same pattern was observed in \cite{kubecon1} as the result of incorrect container labels. 

Our contributions include i) a field failure data analysis (FFDA) of real-world K8s incidents reported in online sources, deriving a failure model for K8s; ii) the design and implementation of \textit{Mutiny}, a K8s injection framework (the data and code are available at \cite{repo}) used to perform a fault/error injection campaign of about 9000 injections in an on-premises cluster; and iii) the comparison of the results of the FFDA and our fault/error injection experiments.

Our key findings are as follows.

\textbf{F1) ${\sim}$3\% (282) of the performed injections led to a system-wide failure.} Those failures include \textit{Stall}, i.e., the cluster's ability to react to changes was compromised, but already-running services remained available; and \textit{Outage}, i.e., services' availability system-wide was compromised. ${\sim}$24\% of injections led to service under/over provisioning, ${\sim}$4\% to service network issues. 
Both experiments and real-world K8s failures show that one incorrect data value can propagate and cause system-wide failures despite the resiliency strategies.

\textbf{F2) Errors in the fields tracking the dependencies among objects caused 51\% of critical failures} (i.e., \textit{Stall}, \textit{Outage}, \textit{Service unreachable}).
Objects are entities representing part of the cluster state. Dependency relationships among them can be dynamically managed through \textit{labels} at the expense of resiliency. Errors in those labels can overload the system, as for the infinite loop of spawned containers cited above \cite{kubecon1}. 

\textbf{F3) Misconfigurations can easily overload the system, like in 13 out of 81 real-world failures.}
The system does not detect hazardous user commands when managing resources at scale. Misconfigurations, which are common in practice\cite{sun2020testing}, can overload the system, and saturate all computing resources.

\textbf{F4) Errors can escape monitoring and propagate inside the system with the user being unaware.}
Orchestrators compare a user-requested desired state (e.g., the number of containers to run) and the observed state (the number of containers spawned). If the observed and desired states differ, the orchestrator takes actions to harmonize (\textit{reconcile}) the two states \cite{burns2016borg}.
When a user modifies the desired state, the orchestrator acknowledges the change but postpones the attainment of the requested state. The observed and desired states should eventually match. If they do not match because of a failure, the user might not receive any error from the system (like in more than 85\% of our experiments), unless proper monitoring alerts are set up. Real-world data show that errors propagated to failures because companies had inappropriate alarms set up.

The implication of the findings is that the system should log changes to labels that can cause critical failures, monitor whether those changes alter system availability, and possibly roll back to the old values when needed.

The novelties of the work include our classification of the orchestration failures, the use of the data store as the target of our injections, and the adoption of our fault/error injection framework for systematic testing of orchestration systems' resiliency as part of the software development process.

%% file: 2_background.tex
\section{Background}
\label{sec:background}

\subsection{Containers and orchestration systems}
\label{subsec:orchestration}
Containers and orchestration systems are technologies that automate the management of services and resources.
Containers provide easy packing of services. 
Orchestrators are distributed systems that provide containers with scheduling, availability, resource allocation, health monitoring, scaling, load-balancing, and networking \cite{khan2017key, rodriguez2019orchestration, burns2016borg}. They are composed of a \textit{control plane} 
and a \textit{compute cluster} made of \textit{worker nodes}.
The former makes global decisions to reconcile the observed state with the desired state received in input.
The latter spawns and monitors the containers assigned to the worker nodes.

\subsection{Putting Mutiny into perspective}

Industry 4.0 \cite{mellado2022design,barletta2023criticality,johansson22kubernetes, barletta2022sla}, 5G networks \cite{nehpio,ferguson2023corekube,botez2021sdn}, avionics \cite{vxworks}, and healthcare \cite{ray2019edge} are a few examples of critical environments in which there are plans to adopt service-based paradigms powered by K8s. In these contexts, a systematic analysis of failure modes is needed.
Among several striking failures reported in blogs \cite{k8sfailures,kubecon1,kubecon2,redditpiday,failuregke,gkefailures}, an interesting example is that of a Reddit compute cluster, in which a node relabeling enforced by a Kubernetes update led to 314 minutes of cluster downtime due to a system-wide network failure \cite{redditpiday}.

As ways to reduce failure-related risks, in several publications \cite{kubecon1,kubecon2} companies mention the use of custom fire-drill tests and \textit{gameday} testing, which deliberately create disruption in K8s to test the orchestrator response and to train cluster operators to do timely troubleshooting. However, these tests differ between companies and do not seem systematic. No established methodology, such as failure mode and effect analysis, field failure data analysis, or Chaos engineering, is mentioned. 
The literature has focused on i) testing of K8s controllers through the discovery of bugs related to distributed system threats, leveraging the injection of probable yet simple faults, e.g., network partitioning, crash, and stale states \cite{sun2022automatic,gu2023acto, lu2019crashtuner, liu2018fcatch, chen2020cofi, sun2021reasoning, majumdar2018random, meng2023distributed}; and ii) assessing, through fault injection, the resiliency of deployed applications  \cite{basiri2016chaos,flora2022study,saurabh400firm}.

Hence, there is a gap in the systematic assessment of the fault tolerance of K8s itself, in terms of component resiliency and recovery capability.
In our preliminary experiments, we applied Chaos engineering to K8s components \revision{with available tools}, but the system was always recovering.
\revision{Indeed, Chaos engineering injects faults/errors (e.g., latencies, crashes, HTTP errors) that are agnostic of the microservices and effective in a complex interaction topology. In K8s, the limited number of components makes simple errors well-tolerated.
Mutiny instead injects state alterations through incorrect values in the data on the datastore.} 
Mutiny can be integrated in the testing clusters \revision{and/or in Chaos engineering processes} to systematically evaluate the response to orchestration errors under realistic workloads and train the cluster operators. Our methods can be extended to custom components, heavily used by companies \cite{kubecon7}, and to other orchestration systems that rely on a main data store containing the whole cluster state and state reconciliation loops \cite{burns2016borg}.

\subsection{Kubernetes} 
\label{subsec:kubernetes}
K8s handles several \textit{resource kinds} (e.g., \textit{Pod}, \textit{ReplicaSet}, and \textit{Node}). For each resource kind, there are multiple \textit{resource instances}. A \textit{Pod} is a set of containers deployed in an isolated environment with some hardware resources allocated. The Pods are stateless or store their states externally (e.g., in \textit{volumes}).
A \textit{ReplicaSet} ensures that a desired number of \textit{Pod replicas} (i.e., Pods running the same application) is running at any one time. 
A \textit{Deployment} manages rolling updates of the container images and the replica number of a ReplicaSet. A \textit{DaemonSet} is similar to a \textit{Deployment}, but it spawns Pods on every \textit{Node} (defined later) that satisfies the constraints. A \textit{Service} provides a single network endpoint to load-balance requests among a set of Pods characterized by a given \textit{Label}. We use the term service for a deployed application that responds to client requests. A Node is either a control plane node or a worker node of the cluster, characterized by a state and available resources.
All resource instances carry some metadata, like \textit{Annotations} and \textit{Labels}, which provide a flexible mechanism to associate resource instances with each other and to define custom information and behaviors.
\begin{figure}[!t]
    \centering
    \includegraphics[width=\columnwidth]{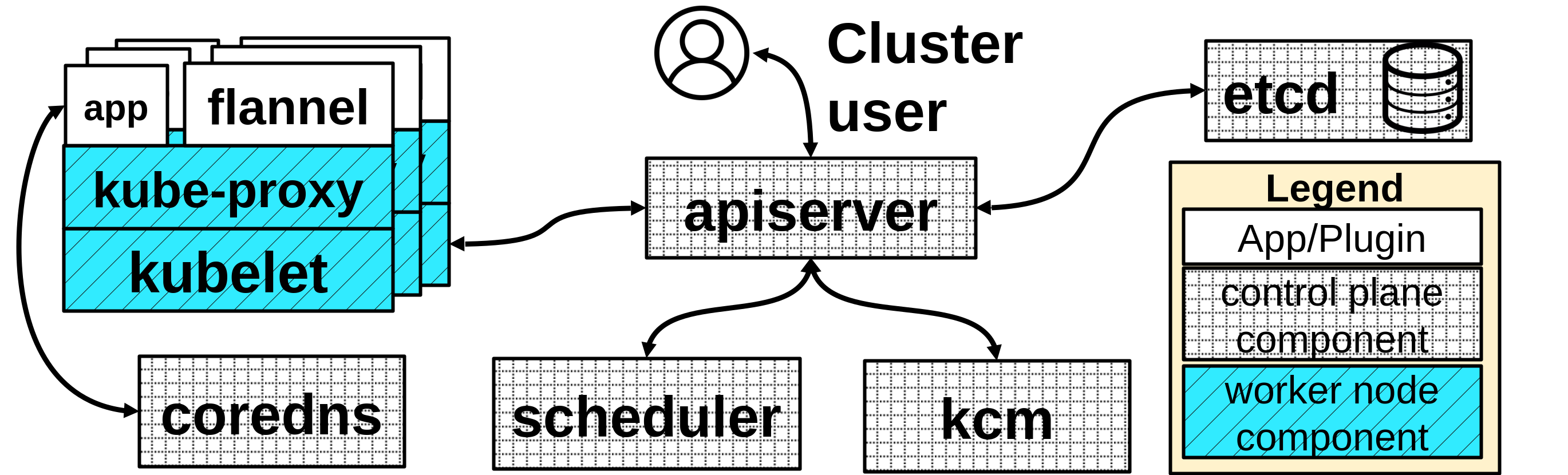}
    \caption{Architecture of K8s.}
    \label{fig:architecture}
    \vspace{-15pt}
\end{figure}

K8s components are loosely coupled 
and include (\figureautorefname~\ref{fig:architecture}) i) \textit{Etcd}, a key-value store guaranteeing sequential consistency to the data of the state of the cluster; ii) \textit{kube-apiserver} (hereafter, \textit{Apiserver}), which exposes the API, allowing users to interact with the cluster and interconnecting the control plane components; iii) \textit{kube-scheduler} (hereafter \textit{Scheduler}), which assigns the Pods to Nodes based on resource requests, availability, and constraints; and iv) \textit{kube-controller-manager} (hereafter, \textit{Kcm}), which reconciles the current cluster state with the desired one. 
The Apiserver communicates with two components deployed on each Node: i) the \textit{kube-proxy}, which maintains the virtual networks, connecting Pods and Services; and ii) the \textit{Kubelet}, which sends Node heartbeats and manages the lifecycle of the assigned Pods, restarting them if they are unhealthy.
By extension, we consider as part of K8s also \textit{coreDNS} and the network manager. They are, respectively, a service providing name translation to the Pods, and a DaemonSet managing networking between Nodes.

The components can be seen as stateless services that indirectly communicate by storing data on Etcd. They hold a cache and can be restarted at any time, fetching necessary data from Etcd. The Apiserver is the only component communicating with Etcd; the others send requests to Apiserver and observe state changes. An event or request can update the state $c_i$ (desired or current) by modifying data on Etcd. All interested components are notified, and they react by creating, deleting, or updating the dependent resource instances, enforcing a linear sequence of state changes, i.e., $[c_i,c_k,...,c_j]$.


\subsection{Resiliency strategies in K8s}
\label{subsec:fault-tolerance}
In the following, we provide a non-exhaustive list of strategies that \revision{K8s uses to increase its resiliency}. 

$\bullet$ Support for optional redundant control plane on different nodes for availability: Etcd, Kcm, and the Scheduler can work in a leader-follower scheme. Etcd uses the Raft consensus algorithm\cite{ongaro2014search} and quorum reads among the replicas. The Kcm and the Scheduler use leader election so that there is only one active replica at a time.

$\bullet$ Level-triggered reconciliation and stateless components: 
decisions are based on the current and desired states\cite{leveltriggered}, and the messages exchanged between components are states, not commands. This guarantees easy recovery if restarts occur. 

$\bullet$ Independent components to foster failure isolation: the control plane components are deployed as independent Pods. A failed component can be restarted by the Kubelet without affecting the other components.

$\bullet$ Circuit breakers that prevent a repeatedly failing operation from overloading the system with a cascading effect. For example, when a Pod fails several consecutive times, it is restarted with increasing back-off delays.

$\bullet$ Timeouts in the communications between components to release resources in a timely manner if failures occur.

$\bullet$ \textit{MaxUnavailability} \revision{to guarantee a minimum number of available replicas during rolling updates of Deployments, limiting the impact of incorrect updates.}

$\bullet$ \textit{Server Side Apply} prevents unauthorized entities from modifying fields of data structures not owned by them. 

$\bullet$ Full disruption mode that stops Pod evictions when all Nodes are reported as unhealthy, since the issue could be in the heartbeat reporting mechanism itself.
    
$\bullet$ Deletion of undecryptable resources \revision{(i.e., resources that cannot be deserialized due to protocol errors)} to prevent a failure during retrieval of the resource lists containing it.


%% file: 3_ffda.tex
\section{Field Failure Data Analysis (FFDA)}
\label{sec:ffda}
This section analyzes data on real-world K8s failures collected in \cite{k8sfailures}. 
The aim is a qualitative analysis to categorize K8s faults, errors, and failures, and to identify fault/error propagation patterns that lead to system/application failures. 
From the analysis it emerges that errors or misconfigurations in subsystems like networking or replication control can cause cluster-wide failures.
A quantitative analysis of the available failure data is not feasible because i) the failures reported (e.g., in online blogs) by companies are a subset (e.g., the most impactful) of all failures that happened in different systems; 
ii) the lack of a systematic approach to data collection may mean failures go unreported; and
iii) the available failure descriptions frequently do not provide relevant failure details and data, hindering a traditional FFDA. 
\revision{Nonetheless, the analysis is useful to analyze and classify the most relevant and impactful error/failure patterns in the wild and inform the design of a fault/error injection campaign.}

\figureautorefname~\ref{fig:fault_error_failure} provides a \textit{Fault-Error-Failure} dependency chain derived from 81 failure instances analyzed.

\begin{table*}[h]
\hspace{-10pt}
\begin{minipage}[t]{0.64\columnwidth}
\begin{adjustbox}{height=111pt}
\begin{tabular}{|p{0.28\textwidth}|p{0.87\textwidth}|p{0.1\textwidth}}
\cline{1-2}
\textbf{Fault} &
\textbf{Fault Description} &
\\ \cline{1-2}
Wrong Autoscale Trigger & Autoscaling of Pods or Nodes is based on misleading information &
\\ \cline{1-2}
Race \newline Condition &
Concurrent actions whose final result depends on timing. E.g., in routing tables/connections &
\\ \cline{1-2}
Unverifiable Certificate &
Certificates cannot be verified or recognized (e.g., cert. rotation)&
\\ \cline{1-2}
Bug & Bug in K8s, third-party, plugins, or underlying code (runtime, OS) &
\\ \cline{1-2}
Human \newline Mistake &
Incorrect command or configuration including: 1) bad resource sizing of components or apps, 2) wrong or badly tuned settings &
\\ \cline{1-2}
Unmanaged Upgrade & System specification or implementation changes, failing regression &
\\ \cline{1-2}
Overload & Too many Pods or Pods with too many resources for a cluster/Node &
\\ \cline{1-2}
Low-Level \newline Issues & Faulty hardware or related drivers &
\\ \cline{1-2}
Failing \newline Application & Misbehaving application causing many events and/or failing Pods &
\\ \cline{1-2}
\multicolumn{2}{c}{(a)}
\end{tabular}%
\end{adjustbox}
\end{minipage}
\hspace{-2pt}
\begin{minipage}[t]{0.63\columnwidth}
\begin{adjustbox}{height=111pt}
\begin{tabular}{|p{0.26\columnwidth}|p{0.77\columnwidth}|p{0.01\columnwidth}}
\cline{1-2}
\textbf{Error} &
\textbf{Error Description} &
\\ \cline{1-2}
State \newline Retrieval &
Irretrievable, stale, or corrupted state due to unavailability, delays or user commands &
\\ \cline{1-2}
Misbehav-ing Logic &
Components behave differently from expected, affecting the reconciliation actions &
\\ \cline{1-2}
Communi- cation &
Networking delays or failures: DNS, routing, load balancing &
\\ \cline{1-2}
Resource \newline Exhaustion &
Affected amount of available computational resources: number of available Nodes, Node/control plane resources, etc. &
\\ \cline{1-2}
Control Plane Availabil. &
Unhealthy control plane components are slowed down or cannot take actions &
\\ \cline{1-2}
Local to \newline worker Nodes &
Errors in underlying software: container runtime, OS, image availability &
\\ \cline{1-2}
\multicolumn{2}{c}{(b)}
\end{tabular}%
\end{adjustbox}
\end{minipage}
\begin{minipage}[t]{0.63\columnwidth}
\begin{adjustbox}{height=111pt}
\begin{tabular}{|p{0.26\columnwidth}|p{1.2\columnwidth}|p{0.1\columnwidth}}
\cline{1-2}
\textbf{Failure} &
\textbf{Failure Description} &
\\ \cline{1-2}
None (\texttt{No}) & System recovered without any consequences, timely reaching the correct steady state &
\\ \cline{1-2}
Timing \newline Failure (\texttt{Tim}) & The creation/update of Pod or other resources took significantly longer than expected, e.g., due to component restarts or overfilled queues.
\\ \cline{1-2}
Less Resources (\texttt{LeR}) & One or a reduced number of services at steady state have permanently allotted less resources than planned, e.g. Pod number or Pod resources &
\\ \cline{1-2}
More Resources (\texttt{MoR}) & One or a reduced number of services has temporarily or permanently allotted more resources than needed, e.g. Pod number or Pod resources &
\\ \cline{1-2}
Service Network (\texttt{Net}) & 
One or a reduced number of services have a correct amount of resources allotted, but incorrectly networked &
\\ \cline{1-2}
Stall \newline (\texttt{Sta}) & Cluster's ability to react to changes was compromised, but already-running services remained unaffected: e.g., new Nodes and Pods not spawned or configured. &
\\ \cline{1-2}
Cluster \newline Outage (\texttt{Out}) & A significant number or all the running services are compromised and unable to respond to application clients anymore &
\\ \cline{1-2}
\multicolumn{2}{c}{(c)}
\end{tabular}%
\end{adjustbox}
\end{minipage}
\caption{Fault-Error-Failure chain of real-world Kubernetes failures. Failures are listed in order of increasing severity.}
\label{fig:fault_error_failure}
\vspace{-5pt}
\end{table*}

\begin{figure*}[!t]
    \centering
    \vspace{-3pt}
    \includegraphics[width=2\columnwidth]{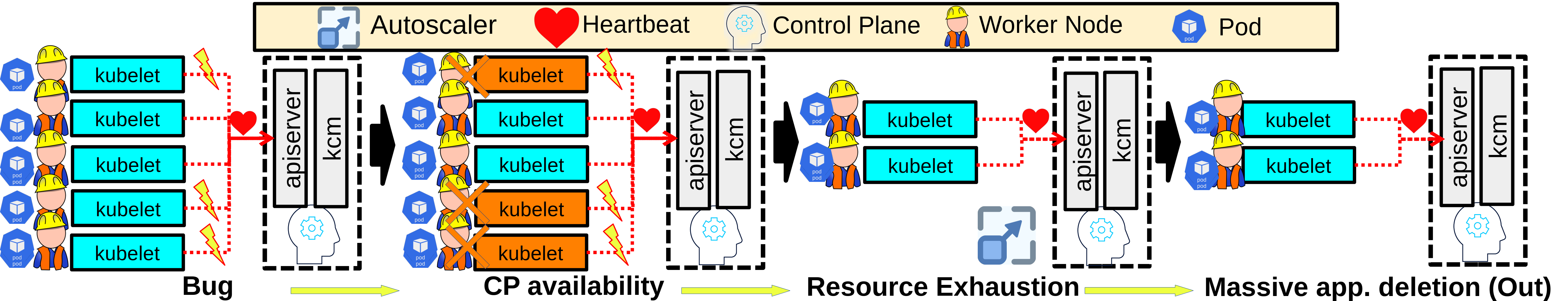}
    \caption{Example cluster outage \texttt{Out} failure. \revision{A timeout during the control plane startup caused an intermittent Apiserver downtime. This caused Kubelets to be unable to report \textit{Node} health, leading to a massive Node deletion and recreation by the Google Kubernetes Engine (GKE) autoscaler.}}
    \label{fig:failureGKE}
    \vspace{-12pt}
\end{figure*}

\subsection{Orchestrator-level failures}
\label{sec:failure_model}
An orchestrator-level failure (OF) is a misbehavior of the orchestration system that may or may not have an impact on applications.
Based on our analysis of the failure data, we classified the K8s failures into the following categories: \textit{Timing Failure} (\texttt{Tim}), \textit{Less Resources} (\texttt{LeR}), \textit{More Resources} (\texttt{MoR}), \textit{Service Network} (\texttt{Net}), \textit{Stall} (\texttt{Sta}), and \textit{Cluster Outage} (\texttt{Out}) (see \figureautorefname~\ref{fig:fault_error_failure} (c)). \revision{Despite being relative to the K8s failure dataset, the failure categories do not depend on any specific feature of K8s and can be used with other orchestration systems as well.}

In our classification, we explicitly differentiate between Cluster Outage (\texttt{Out}) and Stall (\texttt{Sta}) failure types: a Cluster Outage implies that a majority of services are down, while a Stall implies that currently running services are still up, but that the cluster's ability to react to changes (e.g., new user requests or a Node failure) is limited. In an environment with limited evolution, services could remain healthy.
Importantly, error patterns that lead to \texttt{Out} and \texttt{Sta} failures can be similar. For example, spawning an infinite number of Pods can lead to a \texttt{Sta} or \texttt{Out} depending on the Pod priority: preemptive Pods evict all the lower-priority Pods, leading to an \texttt{Out} failure.

We consider the \texttt{MoR} failure type to be more severe than \texttt{LeR} because even if \texttt{LeR} impacts the application SLOs, allotting more resources carries higher costs and risks related to computing resource exhaustion or system overload. 

The differences between \texttt{LeR}, \texttt{Net}, and \texttt{Sta} are mainly in the scale of the failure impact. \texttt{LeR} and \texttt{Net} impact a limited number of services, while \texttt{Out} compromises one of the vital cluster functionalities, impacting almost every running service. For example, a stuck Node might impact a few services, depending on its size, but it does not lead to a system outage. 
\subsection{Orchestrator-level faults and errors}
15 failures in total were \texttt{Out}. Hence, they are not infrequent but a major concern. The most severe faults/errors that caused them had the following causes (in parentheses, the categories from  \figureautorefname~\ref{fig:fault_error_failure}(a,b)): i) network manager failures that impacted the entire cluster (Communication); ii) massive numbers of unhealthy or deleted Nodes (Resource Exhaustion); iii) erroneous commands that deleted namespaces, clusters, or Etcd data (Human Mistake$\rightarrow$State Retrieval); and iv) preemptions caused by infinite spawning (Resource Exhaustion).
For example, \figureautorefname~\ref{fig:failureGKE} illustrates a failure in which a fault hindered the Node heartbeat reporting, leading to massive Node deletion by the Google K8s Engine autoscaler, even if the Nodes were correctly running the applications \cite{failuregke}. 



\subsubsection{Fault/error propagation}\hfill 
\begin{mybox}{F3 - Misconfigurations}
Misconfigurations can easily saturate all computing resources and overload the system, that does not detect hazardous user commands when managing resources at scale.
\end{mybox}
Misconfigurations (Human Mistake in \figureautorefname~\ref{fig:fault_error_failure}(a)) caused 33 of the failures in our data set. 10 of them consisted of bad resource sizing of Nodes and services. 
If services had too few resources, the application failed (Human  Mistake), if they had too many resources, Nodes failed (Overload$\rightarrow$Resource Exhaustion). Specifically, 19 faults were misconfigurations of K8s, 3 misconfigurations of plugins, and 11 misconfigurations of external software.
13 incidents involved errors caused by bugs in K8s code (5), external software (4) (e.g., underlying OS), plugins (1), or custom code (3).
Capacity issues were responsible for 21 failures; 11 of which due to overload of control plane components (Overload, Failing Application, Human Mistake$\rightarrow$Control Plane Availability), which failed to reconcile the cluster state in a timely manner.

19 incidents involved a range of communication errors (Communication in \figureautorefname~\ref{fig:fault_error_failure}(b)): DNS resolution, a misbehaving network manager, blackholes, latencies, and connection errors. They were caused by underlying OS race conditions or bugs, certificate rotations, human mistakes, or unmanaged upgrades. DNS-related issues have been deemed the most painful by multiple companies \cite{kubecon2,kubecon5}.

Various incidents were caused by multiple interacting factors, which are troublesome in conditions rarely met in testing. Often, the alleged root cause of observed failures is a guess, which could be a propagated error of the actual unknown root cause. At other times, it is difficult to derive a cause from the available ``story-telling.'' 
The lack of control motivated us to perform a systematic injection campaign in a controlled environment to better understand the system behavior. 




%% file: 4_methods.tex
\section{Experimental method}
\label{sec:methods}
\begin{figure}[!h]
    \centering
     \includegraphics[width=1\columnwidth]{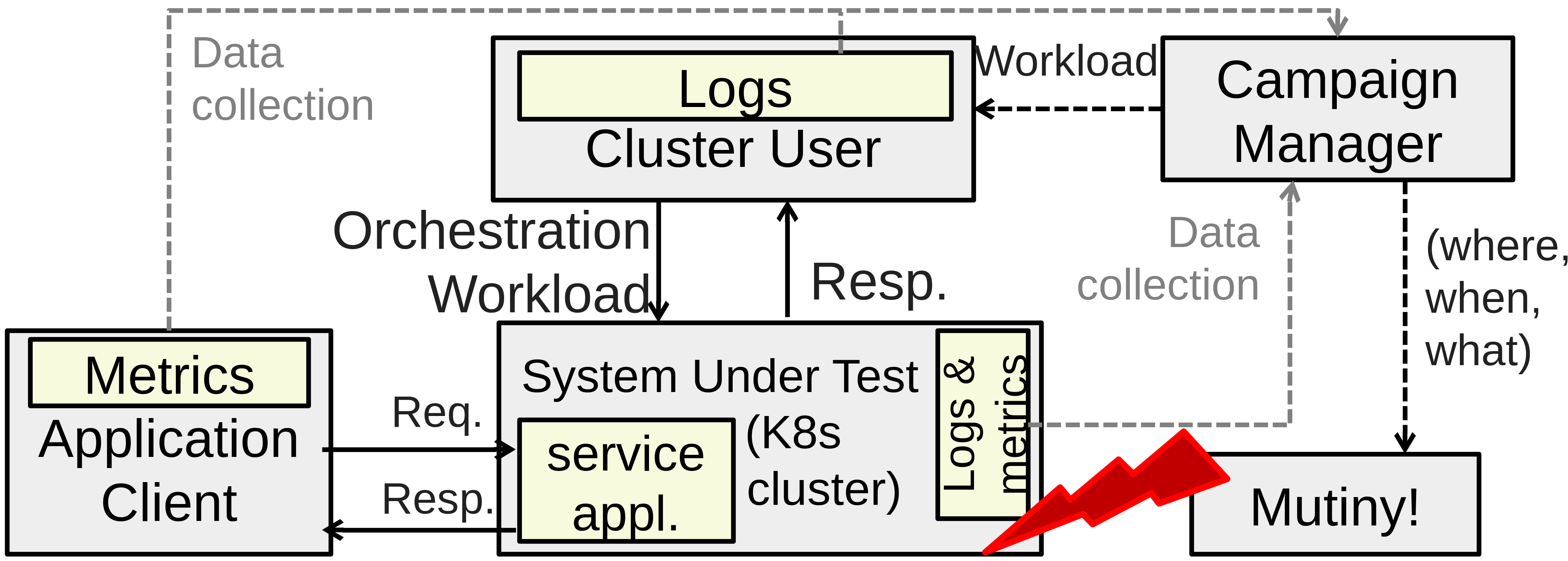}
    \caption{Fault injection framework.}
    \label{fig:exparchitecture}
    \vspace{-5pt}
\end{figure}

This section presents our fault/error injection framework (in \figureautorefname~\ref{fig:exparchitecture}), including workloads, the injector Mutiny, the fault/error injection campaign manager, and the data collection.
\subsection{Mutiny!}
\label{subsec:mutiny}
Mutiny is an injector that can be integrated into K8s to alter the messages exchanged between components and, consequently, the current or desired cluster state. Since K8s components are stateless and level-triggered (\S\ref{sec:background}), all the system state is confined to Etcd, making it a dependability bottleneck 
that can induce failures in any subsystem that stores state information on Etcd. 
We do not care about the possible root causes of alterations: hardware faults, software bugs, misconfigurations, or other causes that somehow store an incorrect value. 
Our fault/error injection framework allows us to systematically inject faults/errors into Etcd and assess system response (in terms of the orchestration actions and application behavior) in a controlled environment. 
\revision{The evaluation is systematic because the framework can introduce faults/errors in all data stored on Etcd under a controlled workload.}
We show (as discussed in \S\ref{sec:results}) that Etcd alterations can recreate a majority ($54/81$) of real-world failures analyzed in \S\ref{sec:ffda}. For example, Nodes can become unhealthy because of a failing Apiserver or a bug in the Kubelet  \cite{failuregke,failureKubelet}. Although the subtle behavior of a bug cannot be replicated, the injections can replicate the effect of having an unhealthy Node, e.g., targeting the heartbeat reporting system.


Three attributes characterize each fault/error injected by Mutiny:
location (where?), type (what?), and trigger (when?).


$\blacksquare$ \textbf{Where} is defined by a communication channel, a resource kind, and either a field value or the serialization protocol bytes of a message. We distinguish two types of communication channels: i) those from Apiserver to Etcd, or ii) those from another component to Apiserver. By injecting the data in the transactions from the Apiserver to Etcd, we directly alter the current or desired cluster state. This emulates faults/errors that originate in the Apiserver or in other components but propagate undetected to Etcd. With replicated control planes, the fault/error is injected before the consensus algorithm is run, so that all Etcd replicas agree on the value. 

Messages directed from other components to the Apiserver undergo authentication, authorization, and admission control. Hence, a corruption of a message in this channel can make the message invalid and cause it to be rejected by the Apiserver. Admission control can change the message content, even through custom code, possibly introducing errors.

$\blacksquare$ \textbf{What} consists of a value and fault/error type between bit-flip, data-type set, and message drop. 

A  bit-flip is an easy way to alter a correct value without understanding its semantics, and hence allows for extensive fault/error injection campaigns. If the value must match regular expressions or ranges, the injected value is, with high probability, still valid but incorrect. Bit-flip faults in Etcd data were also reported by users \cite{issuekbue}. 

A data-type set triggers data validations and integrity checks by setting extreme, invalid, or wrong values, dependent on the field type. Such values might include empty strings, $0$ for integers, or unsupported values for fixed-set values.  

A message drop emulates a state update that did not happen for some reason: a failed request, software bug, updated system specification, or data loss \cite{kubecon3}. It aims to stress the resiliency of level-triggered reconciliation. It is a commonly assumed failure mode in distributed systems \cite{dolev1996failure}. 

$\blacksquare$ \textbf{When} is defined by the occurrence of messages related to the same resource instance sent by the injected component, i.e., the index in the chain $[c^f_i,...,c^f_j]$, where $c^f_{i...j}$ are the state changes (see \S\ref{subsec:kubernetes}) in which the injection target appears. The injection may have different effects depending on the current state of the instance and the next state changes. Moreover, a different occurrence index can correspond to a different action performed by the software, e.g., resource instance creation vs. update. This can influence the transiency of the effect. 

\subsection{Workloads} 
\label{subsec:workloads}

$\blacksquare$ \textbf{The orchestration workloads} 
perform operations on a service application used by a client to create activity in the orchestration system
\footnote{We used synthetic workloads because there is a lack of benchmarks dedicated to the orchestration system. Well-known benchmarks for cloud microservices (e.g., \textit{DeathStarBench} \cite{DeathStartBench}) do not necessarily generate representative orchestration activity (i.e., that activates many orchestration functions).}.
The workloads include
i) \textbf{deploy}, which creates new Deployments and related Pods; ii) \textbf{scale-up}, which increases the replica number of existing Deployments; and iii) \textbf{failover}, in which a Node failure is simulated through a \textit{NoExecution} taint, 
\revision{forcing the Pods running on the Node to be respawned onto available Nodes}. The workloads are applied by \textit{kbench} \cite{kbench} acting as a cluster user (see \figureautorefname~\ref{fig:architecture}). 

$\blacksquare$ \textbf{The service application} is a service exposed to the client. Its characteristics define the orchestration functionalities used.

$\blacksquare$ \textbf{The application client} (AC) sends requests to the service application for a fixed period, monitoring its availability and response times.

\subsection{Campaign manager} 
\begin{figure}[!t]
    \centering
    \includegraphics[width=1\columnwidth]{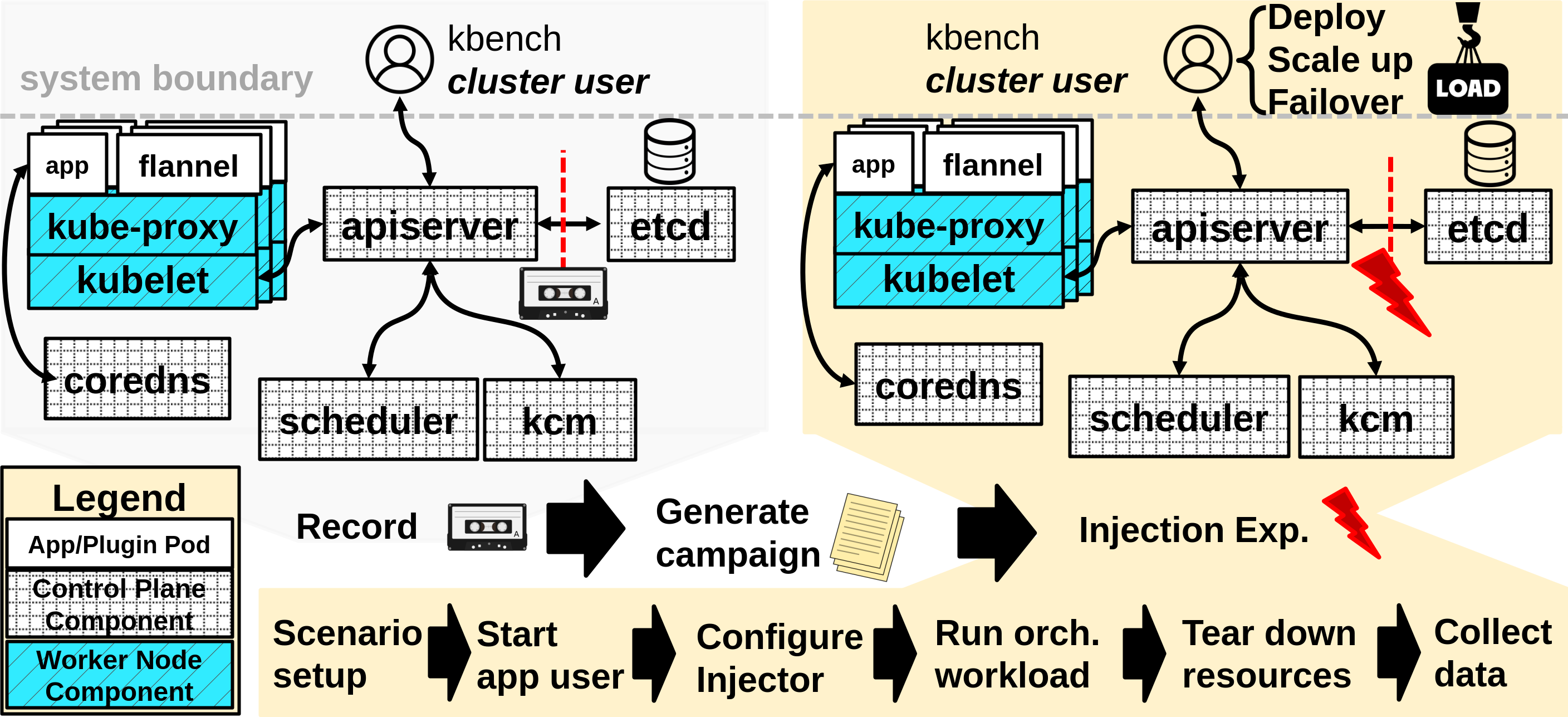}
    \caption{Experimental workflow.}
    \label{fig:approach}
    \vspace{-10pt}
\end{figure}

The campaign manager coordinates fault/error injection experiments, following the workflow in \figureautorefname~\ref{fig:approach}. 
First, we record the fields of the resource instances sent to Etcd during the execution of a nominal orchestration workload, which comprises deploying, scaling, and updating Node states. Later, the injection campaign is generated, and the campaign manager drives the injection experiments.  

$\blacksquare$ \textbf{The injection campaign} includes injection experiments targeting i) a field of a message, ii) its serialization bytes, or iii) a whole message (for message drops). 
For each recorded integer field, we flip a low- and a high-order bit (respectively, $1^{st}$ and $5^{th}$), and we set the $0$ data value. The reason for flipping those bits is that exchanged messages are serialized with the \textit{Protobuf} protocol, and most such encoded integers are long one byte, with the $8^{th}$ bit used as a continuation bit.

For each recorded string field, we flip the least significant bit of the first two characters, and we set the empty string data value. Injecting the least significant bit of a character still results in a character, and hence valid strings.
Boolean fields are inverted.
For each field, we ran an injection experiment for the occurrence indexes $1,2,$ and $3$.
 
For each recorded resource kind, we performed a set of injections targeting random serialization Protobuf bytes to assess the system's response to incorrectly structured messages. 

For each recorded resource kind, we performed a message drop injection for the occurrence indexes from 1 to 10.

After the listed injections, we derived a set of \textit{critical} fields, i.e., fields that caused \texttt{Out}, \texttt{Sta}, or unavailable service failures. We performed additional injection experiments with data-set values specific to the semantics of each critical field.

$\blacksquare$ \textbf{An injection experiment}
\label{subsec:injection_exp}
is composed of the following phases (\figureautorefname~\ref{fig:approach}): K8s cluster restart, fault/error injection scenario set-up, application client workload start, injector programming, orchestration workload execution,  and data collection. In each experiment, a single fault/error was injected.

To restart the cluster, all the Nodes leave the cluster, the control plane Node resets the cluster and creates a new one, and, finally, all the Nodes join the newly created cluster.

The \textit{scenario setup} creates all the resource instances that are required by the orchestration workloads before the injection.

Then, the application client workload starts performing requests to the service application. Next, the campaign manager configures the injection trigger by sending the triplet $(where, when, what)$ in an HTTP request to the injected component. 
\revision{Mutiny is implemented as a package in the K8s source tree. Any component can call it (with instrumentation $<10$ LoC) to tamper a message serialized with Protobuf. 
For bit-flip and data-type set injections, Mutiny de-serializes the message, modifies the content, and re-serializes it, replacing the original.} For message drop injections, the calling function returns without any error before sending the message. In each experiment, we perform a single fault/error injection.

$\blacksquare$ \textbf{The data collection} retrieves the logs of K8s control plane components (with verbosity level set at 6, i.e., debug), Kbench logs, response latencies experienced by the application client, and, finally, the metrics gathered from Prometheus with node\_exporter and kube-state-metrics as sources.

%% file: 5_results.tex
\begin{table}[t]
\begin{minipage}[t]{\columnwidth}
\begin{adjustbox}{height=45pt}
\begin{tabular}{|p{0.31\textwidth}|p{0.885\textwidth}|p{0.01\textwidth}}
\cline{1-2}
\textbf{Failure category} & \textbf{Failure Definition} &  \\ \cline{1-2}
No significant impact (NSI) & The service is available and the response times seen by the AC are not significantly different from golden runs & \\ \cline{1-2}
Higher response times (HRT) & The service is available and the response times seen by the AC are significantly higher from golden runs
& \\ \cline{1-2}
Intermittent availability (IA) & The application client experiences intermittent error responses from the service not due to request timeouts
& \\ \cline{1-2}
Service unreachable (SU) & From a certain instant in time, the service is unreachable to the AC
& \\ \cline{1-2}
\end{tabular}%
\end{adjustbox}
\end{minipage}
\caption{Client failure categories}
\label{tab:CF}
\vspace{-10pt}
\end{table}

\begin{table*}[t!]
\centering
\begin{adjustbox}{width=1.9\columnwidth}
\begin{tabular}{|c|cccc|cccc|cccc|c}
\cline{2-13}
\multicolumn{1}{c|}{} &
\multicolumn{4}{c|}{\textbf{Deploy}} &
\multicolumn{4}{c|}{\textbf{Scale}} &
\multicolumn{4}{c|}{\textbf{Failover}} \\
\cline{2-13}
\multicolumn{1}{c|}{} & NSI & HRT & IA & SU & NSI & HRT & IA & SU & NSI & HRT & IA & SU & \\
\cline{1-13}
\
No & \textbf{1617 (62.2\%)} & \textbf{84 (3.2\%)} & 0 & 0 & \textbf{1382 (54.5\%)} & \textbf{77 (3.0\%)} & 0 & 0 &\textbf{ 2652 (72.7\%)} & \textbf{137 (3.8\%)} & 11 (0.3\%) & 0 & \\
Tim & 28 (1.1\%) & 1 & 0 & 0 & 40 (1.6\%) & 8 & 1 & 0 & 18 (0.5\%) & 11 (0.3\%) & 2 & 0 & \\
LeR & 109 (4.2\%) & 138 (5.3\%) & 4 & 5 & 432 (17.0\%) & 63 (2.5\%) & 0 & 0 & 59 (1.6\%) & 10 (0.3\%) & 1 & 0 & \\
MoR & 368 (14.2\%) & \textbf{12 (0.5\%)} & 2 & 0 & 303 (12.0\%) &\textbf{ 41 (1.6\%)} & 7 & 0 & 531 (14.6\%) & \textbf{31 (0.8\%) }& 0 & 0 & \\
Net & 14 (0.5\%) & 7 & 6 &\textbf{ 107 (4.1\%)} & 28 (1.1\%) & 46 (1.8\%) & 10 (0.4\%) & 0 & 8 & 48 (1.3\%) & \textbf{40 (1.1\%)} & 1 & \\
Sta &\textbf{ 81 (3.1\%)} & \textbf{4} & 0 & 0 & \textbf{81 (3.2\%)} & \textbf{5} & 0 & 0 & \textbf{66 (1.8\%)} & \textbf{8} & 0 & 0 & \\
Out &\textbf{ 10 (0.4\%)} & \textbf{1} & 0 & \textbf{1} & \textbf{8} & \textbf{1} & 0 & \textbf{1} & \textbf{7} & \textbf{2} & 2 & \textbf{4} & \\

\cline{1-13}
\end{tabular}
\end{adjustbox}
\caption{Mapping between orchestrator failures (OF) and client failures (CF). Percentages are of the total number of injections performed for that given workload. Percentages of single-digit numbers are omitted for readability.}
\label{tab:OFCFmapping}
\vspace{-10pt}
\end{table*}

\section{Experimental results}
\label{sec:results}
In this section, we describe the results of the fault/error injection campaign. The experiments' aims were to understand K8s's resiliency to faults/errors, pinpoint the inherent weaknesses of K8s that can trigger severe failures, and characterize the impact of orchestration failures on services.
\subsection{Experimental setup and parameters}
Our experimental setup consisted of a cluster running K8s v1.27.4 in the default \textit{kubeadm} configuration. 
The cluster included 1 control plane Node and 4 worker Nodes, one of which was used for the application client and monitoring Pods The network manager was flannel v1.1.2. 
\revision{The cluster featured the default resiliency strategies described in \S\ref{subsec:fault-tolerance}. Unless differently specified (see \S\ref{subsec:failuresOFCF}) the cluster was managed by a single control plane Node running all control plane Pods (default configuration\cite{components}). Although production environments commonly use multiple control plane Nodes, this improves the availability in case of a Node crash but provides no protection from faulty values on the datastore. Indeed, the Kcm and Scheduler have only one active replica at any time, while the datastore replicas agree on the faulty value.} 
The nodes were virtual machines (8 CPU, 4 GB RAM, \revision{Ubuntu 20.04, Linux kernel v5.4, containerd v1.7}) communicating through an internal network, on top of VirtualBox 6 hypervisor in a cloud-tier environment (Intel Xeon E5-2695), where no other user application was running.

Based on the amount of resources of our setup, we parametrized the workloads as follows: the ``deploy" workload created three Deployments, each with two replicas; the ``scale-up" workload scaled two \revision{Deployments from two replicas each, to three replicas each, after 10 seconds to four each}, and after another 10 seconds to five each; and the ``failover" workload deals with three running Deployments with two replicas each. Kbench waited up to 40 seconds for each request to be completed.
The service application was a \textit{Flask} webserver, \revision{which read a seed for random numbers from a \textit{Volume} during the startup, and responded to clients with the result of random computations.} 
Its Pods had CPU and memory resource requests and limits, and default priority.
The web server was stateless and did not require coreDNS name resolution. \revision{The application is used to trigger orchestration activity, through the workloads defined in \S\ref{subsec:workloads}: a stateful application with complex topology would complicate the application failure patterns but not the orchestration system ones.} 
The application client sent 20 requests/second for 30 seconds.

\subsection{Analysis of data from fault/error injection campaign}
We consider two levels of failures: orchestration-level failures (OF, \S\ref{sec:ffda}), and client-level failures (CF). CFs reveal the fault/error impact on application clients (AC) in terms of performance and availability.
For both OFs and CFs, if a failure belonged to more than one category, we classified it as the most severe failure category.

In \tableautorefname~\ref{tab:CF}, we introduce the categories of client failures: no significant impact (NSI), higher response times (HRT), intermittent availability (IA), service unreachable (SU). 
For each workload, we collected data from 100 golden runs without any faults/errors injected.

$\blacksquare$ \textbf{To classify orchestration-level failures}, for every golden run we collected the number of ready replicas for each ReplicaSet, and the number of Service endpoints, every 3 seconds. We collected Kbench statistics regarding the number of Pods created/scheduled/running and the Pods' \textit{total startup times}, as defined in \cite{kbench}. We classify the failures as follows, recalling \figureautorefname~\ref{fig:fault_error_failure} (c).

\texttt{Tim} failure: A service Pod is restarted, or the z-score relative to the golden distribution of either the worst Pod total startup time or the last Pod creation time is greater than 3.

\texttt{LeR} failure: The number of ready replicas, created Pods, or endpoints is stable and lower than the baseline. 

\texttt{MoR} failure: The number of ready replicas, created Pods, or endpoints is higher than the baseline. 

\texttt{Net} failure: The number of ready replicas and Pods is correct, but some are not reachable or used in load-balancing.

\texttt{Sta} failure: There is an uncontrolled Pod spawn, control plane Pods are stuck, or networking Pods fail.

\texttt{Out} failure: All the ReplicaSets are unreachable (including Prometheus), the DNS Pods fail, or the networking Pods fail and cause a disruption of the service application.

\begin{figure}[!b]
\vspace{-15pt}
 \centering
 \includegraphics[width=\columnwidth]{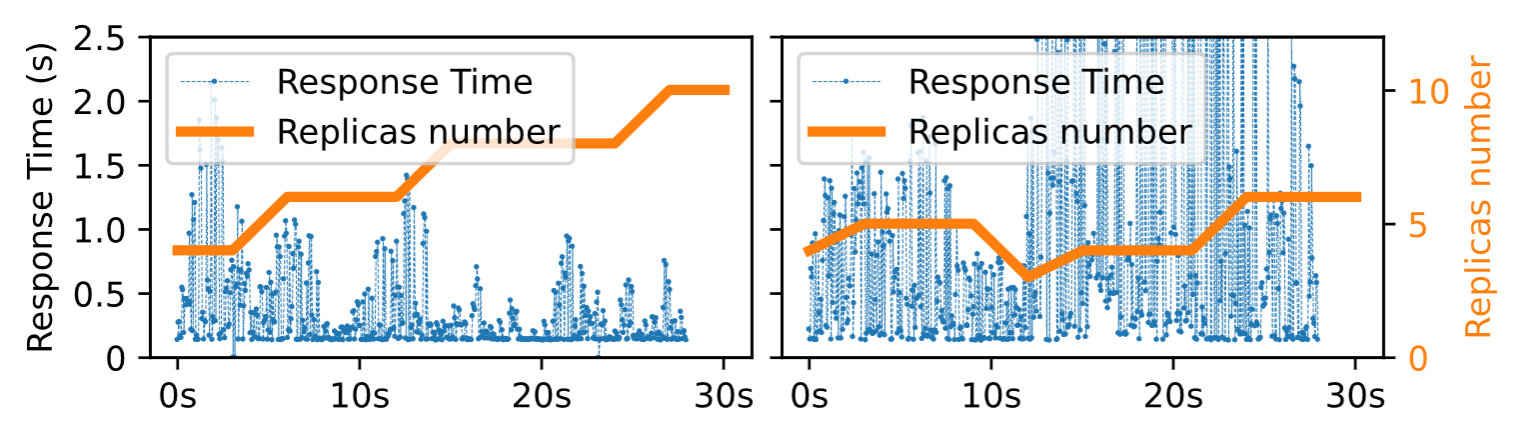}
 \caption{On the left, a golden run time series ($z\_score=-0.2$). On the right, an injection time series ($z\_score=11.0$)}
 \label{fig:profileexample}
\end{figure}

$\blacksquare$ \textbf{To classify client-level failures}, \revision{we create a time series for each golden run containing the response time latency of the requests, ordered by the time of sending}. We padded with 0 the response times of failed requests. We computed a \textit{baseline} time series for each workload by averaging the golden run time series. We measured the Mean Absolute Error (MAE) between each golden run time series and the baseline time series, obtaining a distribution of golden-run MAEs.
For each injection experiment, we computed the MAE between the experimental time series and the baseline, and computed the z-score of the MAE against the distribution of golden-run MAEs. The z-score quantifies the impact on the application client (see \figureautorefname~\ref{fig:profileexample}). We classify the failures as follows.

\texttt{HRT} failure: The z-score is greater than 2. 

\texttt{IA} failure: The application client experiences intermittent errors not due to timeouts.

\texttt{SU} failure: An application has no response from the service. 

\subsection{Results}
\label{subsec:results}
We performed a total of  8,782 injection experiments based on the campaign described in \S\ref{subsec:injection_exp}, targeting the communication between the Apiserver and Etcd \revision{in order to directly alter the stored state and efficiently trigger failures, as mentioned in \S\ref{subsec:mutiny}}. Exactly one fault/error was injected in each experiment.

The results are summarized in \tableautorefname s~\ref{tab:OFCFmapping},  ~\ref{tab:client_and_state}, and \ref{tab:client_failures}. 
\tableautorefname~\ref{tab:OFCFmapping} describes the propagation of OFs to CFs. For example, the cell intersecting the column HRT and row \texttt{MoR} contains the number of \texttt{MoR} failures that caused HRT. \tableautorefname s~\ref{tab:client_and_state}, and \ref{tab:client_failures} divide failures by workload and injection type, with percentages of categories, e.g., $2.8\%$ of experiments are \texttt{Sta}. 

\begin{table}[h!]
\resizebox{\columnwidth}{!}{
\begin{tabular}{|l|l|l|lllllll|}
\cline{1-10}
\multirow{2}{*}{\textbf{WL}} &
 \multirow{2}{*}{\textbf{Injection}} &
 \multirow{2}{*}{\textbf{Perf.}} &
 \multicolumn{7}{c|}{\textbf{Orchestration-level Failures (OF)}}
 \\ \cline{4-10}
 &
 &
 &
 \multicolumn{1}{l|}{No} &
 \multicolumn{1}{l|}{Tim} &
 \multicolumn{1}{l|}{LeR} &
 \multicolumn{1}{l|}{MoR} &
 \multicolumn{1}{l|}{Net} &
 \multicolumn{1}{l|}{Sta} &
 \multicolumn{1}{l|}{Out}
 \\ \cline{1-10}
\rowcolor{Gray}
\cellcolor{white}
\multirow{3}{*}{Deploy} &
Bit-flip &
1563 &
\multicolumn{1}{l|}{1097} &
\multicolumn{1}{l|}{17} &
\multicolumn{1}{l|}{135} &
\multicolumn{1}{l|}{210} &
\multicolumn{1}{l|}{58} &
\multicolumn{1}{l|}{45} &
\multicolumn{1}{l|}{1}
\\ \cline{2-10}
&
Value set &
900 &
\multicolumn{1}{l|}{484} &
\multicolumn{1}{l|}{12} &
\multicolumn{1}{l|}{111} &
\multicolumn{1}{l|}{172} &
\multicolumn{1}{l|}{70} &
\multicolumn{1}{l|}{40} &
\multicolumn{1}{l|}{11}
\\ \cline{2-10}
\rowcolor{Gray}
\cellcolor{white}
&
Drop &
136 &
\multicolumn{1}{l|}{120} &
\multicolumn{1}{l|}{0} &
\multicolumn{1}{l|}{10} &
\multicolumn{1}{l|}{0} &
\multicolumn{1}{l|}{6} &
\multicolumn{1}{l|}{0} &
\multicolumn{1}{l|}{0}
\\ \cline{1-10}
&
Bit-flip &
1522 &
\multicolumn{1}{l|}{950} &
\multicolumn{1}{l|}{29} &
\multicolumn{1}{l|}{260} &
\multicolumn{1}{l|}{190} &
\multicolumn{1}{l|}{45} &
\multicolumn{1}{l|}{47} &
\multicolumn{1}{l|}{1}
\\ \cline{2-10}
\rowcolor{Gray}
\cellcolor{white}
&
Value set  &
872 &
\multicolumn{1}{l|}{387} &
\multicolumn{1}{l|}{17} &
\multicolumn{1}{l|}{224} &
\multicolumn{1}{l|}{161} &
\multicolumn{1}{l|}{35} &
\multicolumn{1}{l|}{39} &
\multicolumn{1}{l|}{9}
\\ \cline{2-10}
\multirow{-3}{*}{Scale} &
Drop &
140 &
\multicolumn{1}{l|}{122} &
\multicolumn{1}{l|}{3} &
\multicolumn{1}{l|}{11} &
\multicolumn{1}{l|}{0} &
\multicolumn{1}{l|}{4} &
\multicolumn{1}{l|}{0} &
\multicolumn{1}{l|}{0}
\\ \cline{1-10}
\rowcolor{Gray}
\cellcolor{white}
\multirow{3}{*}{Failover} &
Bit-flip &
2132 &
\multicolumn{1}{l|}{1610} &
\multicolumn{1}{l|}{13} &
\multicolumn{1}{l|}{5} &
\multicolumn{1}{l|}{424} &
\multicolumn{1}{l|}{33} &
\multicolumn{1}{l|}{42} &
\multicolumn{1}{l|}{5}
\\ \cline{2-10}
&
Value set  &
1288 &
\multicolumn{1}{l|}{972} &
\multicolumn{1}{l|}{18} &
\multicolumn{1}{l|}{64} &
\multicolumn{1}{l|}{130} &
\multicolumn{1}{l|}{62} &
\multicolumn{1}{l|}{32} &
\multicolumn{1}{l|}{10}
\\ \cline{2-10}
\rowcolor{Gray}
\cellcolor{white}
&
Drop &
229 &
\multicolumn{1}{l|}{218} &
\multicolumn{1}{l|}{0} &
\multicolumn{1}{l|}{1} &
\multicolumn{1}{l|}{8} &
\multicolumn{1}{l|}{2} &
\multicolumn{1}{l|}{0} &
\multicolumn{1}{l|}{0}
\\ \cline{1-10}
\cline{4-10}
\multicolumn{1}{l}{} &
\multicolumn{1}{l}{$\sum$} &
\multicolumn{1}{l|}{8782} &
\multicolumn{1}{l|}{5960} &
\multicolumn{1}{l|}{109} &
\multicolumn{1}{l|}{821} &
\multicolumn{1}{l|}{1295} &
\multicolumn{1}{l|}{315} &
\multicolumn{1}{l|}{245} &
\multicolumn{1}{l|}{37}
\\
\multicolumn{1}{l}{} &
\multicolumn{1}{l}{\%} &
\multicolumn{1}{l|}{100\%} &
\multicolumn{1}{l|}{\textbf{67.8\%}} &
\multicolumn{1}{l|}{1.2\%} &
\multicolumn{1}{l|}{9.4\%} &
\multicolumn{1}{l|}{14.8\%} &
\multicolumn{1}{l|}{3.6\%} &
\multicolumn{1}{l|}{\textbf{2.8\%}} &
\multicolumn{1}{l|}{\textbf{0.4\%}}
\\ \cline{4-10}
\end{tabular}%
}
\caption{Statistics on orchestrator-level (OF) failures observed in fault/error injection experiments.}
\label{tab:client_and_state}
\vspace{-5pt}
\end{table}

\begin{table}[h]
\centering
\resizebox{0.75\columnwidth}{!}{%
\begin{tabular}{|l|l|l|llll|}
\cline{1-7}
\multirow{2}{*}{\textbf{WL}} &
 \multirow{2}{*}{\textbf{Injection}} &
 \multirow{2}{*}{\textbf{Perf.}} &
 \multicolumn{4}{c|}{\textbf{Client-level Failures (CF)}}
 \\ \cline{4-7}
 &
 &
 &
\multicolumn{1}{l|}{NSI} &
\multicolumn{1}{l|}{HRT} &
\multicolumn{1}{l|}{IA} &
 SU
 \\ \cline{1-7}
\rowcolor{Gray}
\cellcolor{white}
\multirow{3}{*}{Deploy} &
Bit-flip &
1563 &
\multicolumn{1}{l|}{1386} &
\multicolumn{1}{l|}{132} &
\multicolumn{1}{l|}{5} &
\multicolumn{1}{l|}{40}
\\ \cline{2-7}
&
Value set &
900 &
\multicolumn{1}{l|}{720} &
\multicolumn{1}{l|}{105} &
\multicolumn{1}{l|}{6} &
\multicolumn{1}{l|}{69}
\\ \cline{2-7}
\rowcolor{Gray}
\cellcolor{white}
&
Drop &
136 &
\multicolumn{1}{l|}{121} &
\multicolumn{1}{l|}{10} &
\multicolumn{1}{l|}{1} &
\multicolumn{1}{l|}{4}
\\ \cline{1-7}
&
Bit-flip &
1522 &
\multicolumn{1}{l|}{1379} &
\multicolumn{1}{l|}{133} &
\multicolumn{1}{l|}{10} &
\multicolumn{1}{l|}{0}
\\ \cline{2-7}
\rowcolor{Gray}
\cellcolor{white}
&
Value set &
872 &
\multicolumn{1}{l|}{772} &
\multicolumn{1}{l|}{91} &
\multicolumn{1}{l|}{8} &
\multicolumn{1}{l|}{1}
\\ \cline{2-7}
\multirow{-3}{*}{Scale} &
Drop &
140 &
\multicolumn{1}{l|}{123} &
\multicolumn{1}{l|}{17} &
\multicolumn{1}{l|}{0} &
\multicolumn{1}{l|}{0}
\\ \cline{1-7}
\rowcolor{Gray}
\cellcolor{white}
\multirow{3}{*}{Failover} &
Bit-flip &
2132 &
\multicolumn{1}{l|}{1989} &
\multicolumn{1}{l|}{132} &
\multicolumn{1}{l|}{9} &
\multicolumn{1}{l|}{2}
\\ \cline{2-7}
&
Value set &
1288 &
\multicolumn{1}{l|}{1139} &
\multicolumn{1}{l|}{100} &
\multicolumn{1}{l|}{46} &
\multicolumn{1}{l|}{3}
\\ \cline{2-7}
\rowcolor{Gray}
\cellcolor{white}
&
Drop &
229 &
\multicolumn{1}{l|}{213} &
\multicolumn{1}{l|}{15} &
\multicolumn{1}{l|}{1} &
\multicolumn{1}{l|}{0}
\\ \cline{1-7}
\cline{4-7}
\multicolumn{1}{l}{} &
\multicolumn{1}{l}{$\sum$} &
\multicolumn{1}{l|}{8782} &
\multicolumn{1}{l|}{7842} &
\multicolumn{1}{l|}{735} &
\multicolumn{1}{l|}{86} &
\multicolumn{1}{l|}{119}
\\
\multicolumn{1}{l}{} &
\multicolumn{1}{l}{\%} &
\multicolumn{1}{l|}{100\%} &
\multicolumn{1}{l|}{89.2\%} &
\multicolumn{1}{l|}{\textbf{8.4\%}} &
\multicolumn{1}{l|}{\textbf{0.9\%}} &
\multicolumn{1}{l|}{\textbf{1.4\%}}
\\ \cline{4-7}
\end{tabular} }
\caption{Statistics on client-level (CF) failures observed in fault/error injection experiments.}
\label{tab:client_failures}
\vspace{-10pt}
\end{table}

\subsubsection{Analysis of OF and CF failures}\label{subsec:failuresOFCF} \hfill
\begin{mybox}{\textbf{F1} - System-wide failures}
$3.2$\% of the performed injections of one value propagated to a system-wide failure,  despite the resiliency strategies. 
$24.2$\% of injections resulted in service under/over provisioning, $3.6$\% in service networking problems.
${\sim}70\%$ of performed injections have no effect because they are either i) detected and mitigated by the health checks, like heartbeats; or ii) mitigated by natural system behavior (e.g., the value is overwritten). 
\end{mybox}

In our experiments, a non-negligible number ($3.2\%$, last two columns of \tableautorefname~\ref{tab:client_and_state}) of fault/error injections of a single bit-flip or value set resulted in \texttt{Sta} and \texttt{Out} failures. \texttt{Sta} failures were caused by i) a control plane overload due to uncontrolled replication of resource instances (e.g., Pods); ii) a Scheduler or Kcm that was unable to obtain a leadership role and perform state changes; or iii) a failure or deletion of networking Pods. 
On the other hand, the causes of cluster outages were i) uncontrolled replication of resource instances; ii) misconfigured networking daemons that caused a global network outage; or iii) coreDNS Pods that failed or were deleted. Below there is an example of uncontrolled replication.

\begin{table}[htbp]
    \centering
    \begin{tabular}{|p{0.97\columnwidth}|} 
    \hline
    \rowcolor{Gray}
    \revision{Example of uncontrolled replication} \\
    \hline
A single-bit corruption of the labels that associate a Pod with a DaemonSet leaves the Kcm unable to identify the Pods belonging to the DaemonSet. That causes new Pods to be spawned, in an infinite loop. The system is overloaded and all the cluster computing resources are filled up. The DaemonSet Pods have high scheduling priority, so they terminate all application Pods to claim resources. 
Eventually, the disk of the control plane Node can fill up, stalling Etcd. \\
    \hline
    \end{tabular}
\end{table}


Particularly interesting is the case of injections affecting the serialization protocol. They usually cause the resource instance to become undecryptable and be deleted (see \S\ref{sec:background}), but in some cases, the resource instance remains decryptable and wrong. Because of how the protocol works, an injection can move a value from one field to another, and a required field could remain empty and trigger failures.


Our results indicate that $3.6\%$ of injections resulted in service networking problems (column 5 in \tableautorefname~\ref{tab:client_and_state}), $24.2\%$ in service under/over provisioning (column 3,4).

${\sim} 70\%$ of faults/errors across the three workloads had no perceivable effect (first column in \tableautorefname~\ref{tab:client_and_state}). 
Both the injections recovered and the ones not activated belonged to this set. 
We define an injection as activated when the injected resource instance is requested after the injection. The activation rate is 82\%. We have no control over the activation of a single field.



Examples of system recovery include 
i) overwriting of the injected data field with a correct value that is still stored somewhere in the system (e.g., some ReplicaSet fields\revision{, which cause a ReplicaSet recreation} or the \textit{PodIP}\revision{, which is overwritten by the correct value sent by kubelets}) or 
ii) the corrupted data have no immediate effect but remain latent. \revision{For example, some data-structures have a versioning number. If the Kcm does not detect any change in the number, it does not process the instance, preventing the injected value from being used}. However, a subsequent update of the versioning number (e.g., by another request) triggers the errors caused by the injected value. For example, several injections targeting the networking DaemonSets can lead to a \texttt{Sta} or \texttt{Out} if triggered.

Injections classified as \texttt{No} mostly did not propagate to clients (see \texttt{No}-NSI cell in \tableautorefname~\ref{tab:OFCFmapping}). 
However, some of them led to \texttt{HRT} client failure. Those cases could be attributed to the natural nondeterministic timing behavior of the orchestrator.


\begin{figure}[!b]
\vspace{-20pt}
 \centering
 \includegraphics[width=\columnwidth]{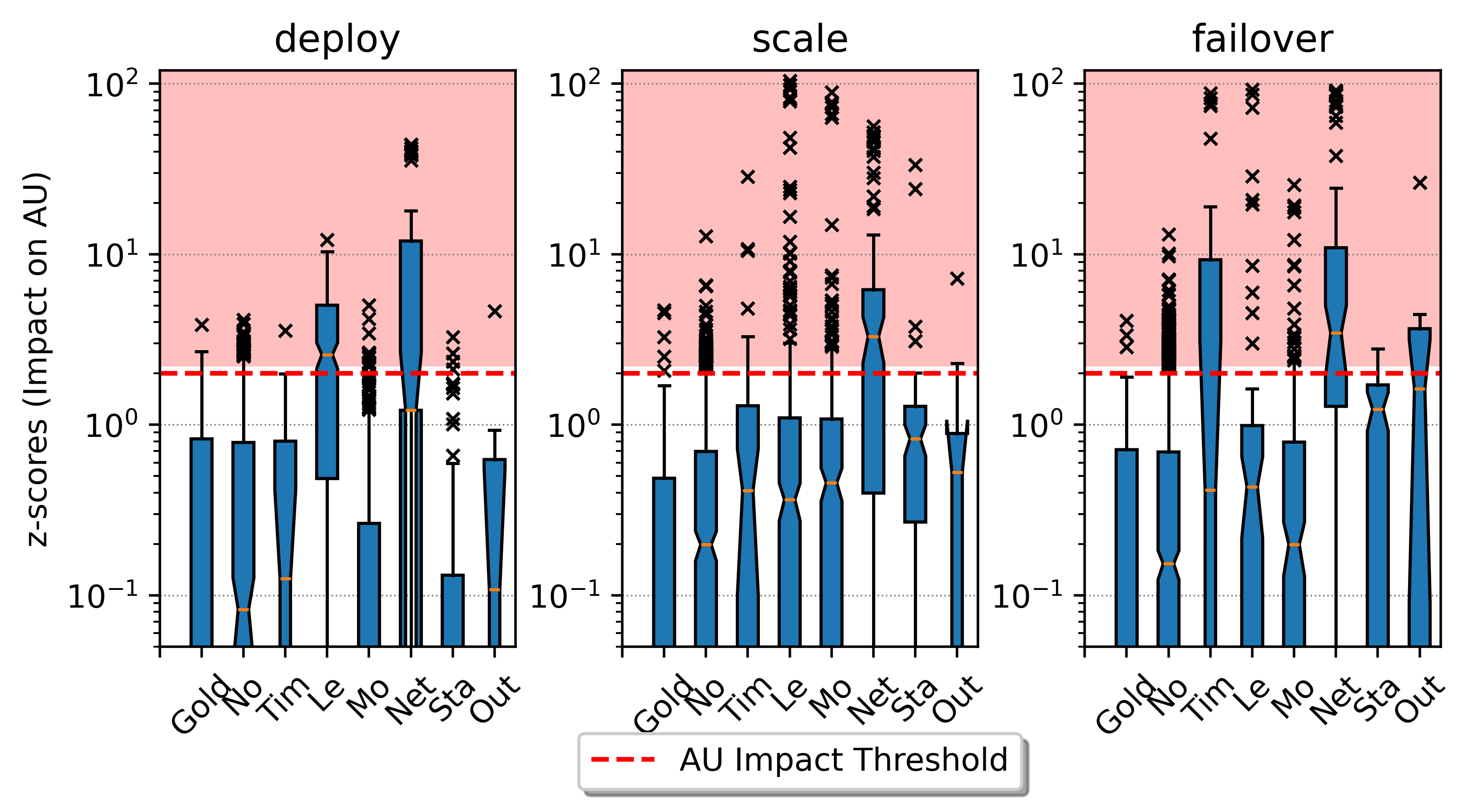}
 \caption{Impact on app. client measured through z-scores.}
 \label{fig:zscores}
\end{figure}

A non-negligible number ($10.8\%$, last three columns of \tableautorefname~\ref{tab:client_failures}) of injections impact the clients. \figureautorefname~\ref{fig:zscores} shows the z-scores of response times observed by the clients for different fault/error types injected at the orchestrator level. 

Timing failures generally have a limited impact on clients, but under the ``failover" workload, they can introduce a significant delay (reflected by a high z-score) because of control plane Pods' restarts.
Scheduler restarts were caused by injections into the \textit{nodeName} field of an existing Pod, which changed the value to a non-existing \textit{nodeName}. 

Below, an example of this phenomenon.
\vspace{-5pt}
\begin{table}[htbp]
    \centering
    \begin{tabular}{|p{0.97\columnwidth}|} 
    \hline
    \rowcolor{Gray}
    \revision{Example of timing failure} \\
    \hline
The Scheduler detects a mismatch between the data in Etcd and its cache and, assuming a cache corruption, restarts.
After a new leader Scheduler is elected (after 20 seconds, in the standard configuration), it starts scheduling Pods. The corrupted Pod remains pending for ${\sim}50$ seconds, until the Kcm deletes it and creates another one. \\
    \hline
    \end{tabular}
\end{table}
\vspace{-5pt}

Less resources failures can have severe impacts on response times when the difference between the expected and used numbers of Pods is significant. 
We observed that part (${\sim}40\%$) of \texttt{MoR} failures represent a negligible threat to clients because they are transient and involve little extra resource consumption. (The Pod number at steady state is correct, but the number of Pods spawned is greater than expected by less than three.)
Interestingly, \texttt{MoR} failures can negatively impact the clients as well. 
When the system does not detect the resource overprovisioning for a service and uses less resources than allotted, a \texttt{LeR} and \texttt{MoR} failures are caused at the same time.
\vspace{-10pt}
\begin{table}[htbp]
    \centering
    \begin{tabular}{|p{0.97\columnwidth}|} 
    \hline
    \rowcolor{Gray}
   \revision{ Example of undetected overprovisioning }\\
    \hline
When the \textit{namespace} field of a Deployment is corrupted during the scale-up of the service application, fewer Pods are spawned, causing longer response times. Upon deletion of the resource instances, K8s starts reconciling a residual corrupted Deployment that is not even listed anymore, spawning Pods in the \textit{terminating} state in an infinite loop. When the rates of terminated and created Pods become similar, the system reaches an equilibrium, but Etcd is filling up. \\
    \hline
    \end{tabular}
\end{table}
\vspace{-5pt}

Service networking failures \texttt{(Net)} induce the majority of intermittent failures for clients (\texttt{IA}), and complete service outages (\texttt{SU}). Almost all \texttt{SU}s happen under deploy (see \tableautorefname~\ref{tab:OFCFmapping}); the injections with index 1 during deploy are create transactions, making the unwanted value changes barely detectable, unlike the injections in the following updates.
\texttt{Sta} failures may or may not impact the application client response times, as said in \S\ref{sec:failure_model}, although the system eventually gets to a degraded state. Finally, \texttt{Out} failures in our data do not always impact response times because the service application does not require the \textit{DNS}. This makes error propagation from the orchestrator to the clients difficult to identify.

We repeated the injections targeting the \textit{critical} data fields (360 in total; see \S\ref{paragraph:criticalfield}) in a cluster with three control plane Nodes with an Etcd replica on each control plane Node.
The results show no significant difference from the previous ones. The only component that actually works in a replicated fashion is Etcd, and values were injected before getting to it. 
A few additional experiments also showed that corrupting the data in Etcd at rest has a different propagation pattern from our injections because of the Apiserver cache. The cache is used intensively, and it is refreshed with Etcd data when needed, e.g., getting
a resource instance. If the refresh does not happen before an update, the injected value in Etcd is overwritten, and a complete component restart may be needed to pick up the injected value. Furthermore, quorum reads mitigate corrupted values.
In conclusion, i) corruption at rest is less likely to cause issues than errors that happen before a transaction; and ii) a corruption of the cache may overwrite a correct value on the data store if the right sequence of requests is triggered.

\subsubsection{Critical field analysis}\hfill
\label{paragraph:criticalfield}
\begin{mybox}{\textbf{F2} - Dependency relationships}
$51\%$ of fault/error injection experiments that caused critical failures targeted the fields managing the dependency relationships among resource instances, revealing an inherent data weakness.
\end{mybox}

\noindent We analyzed the fields that caused the most severe failures when injected, i.e., \texttt{Sta}, \texttt{Out}, or \texttt{SU}. $377$ injections were derived, which all affected the same $34$ fields of different resource kinds. Out of them, $8$ are related to metadata and $26$ to technical specifications. 

A subset of fields is important because it constitutes the way in which K8s keeps track of the associations between multiple dependent resource instances of different kinds. These include: i) owner relationships with references to other resource instances, and ii) label relationships that use matching labels and selectors to create dynamic relationships. \textit{labels}, \textit{managed-by}, \textit{targetRef}, and \textit{ownerReferences} are metadata, while label selectors are specification fields.
In total, $20$ fields out of $34$ belong to this subset, representing $187$ injections out of $360$.
The injections that triggered uncontrolled replication of objects belong to this category, revealing an issue in K8s: although these fields are important for the functioning and represent a risk because of the associated possible failures, there are not enough resiliency strategies in place to recover the system in case of errors. 
Other relevant fields ($124$ injections total) are \textit{name}, \textit{namespace}, and \textit{uid}, which are the fields used by K8s to identify a resource and appear in its URL. The remaining fields include $5$ related to networking (protocols, addresses, and ports); the replica number; and $2$ specification fields of images and commands that prevent the start of critical Pods.


\begin{figure}[!t]
\centering
\begin{subfigure}{\columnwidth}
 \includegraphics[width=1\columnwidth]{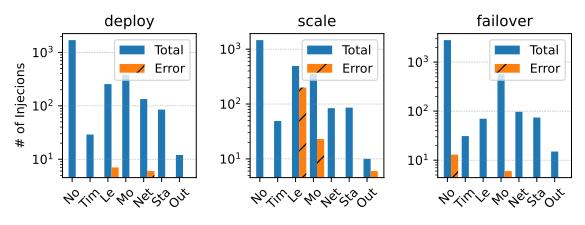}
\end{subfigure}
 \caption{Number of total injection experiments vs. injection experiments in which the cluster user received an error (\textit{Error}) in response to requests to the Apiserver.}
  \label{fig:APIerror}
\vspace{-15pt}
\end{figure}

\subsubsection{User error analysis}\hfill
\begin{mybox}{\textbf{F4} - User unawareness}
The reconciliation of the observed cluster state and the desired state is postponed to a later time. If the state of the system diverges and never reaches the desired state because of failures, the user may be unaware of it unless proper monitoring alerts are set.
\end{mybox}

\figureautorefname~\ref{fig:APIerror} shows the number of injection experiments with failed user requests to the Apiserver (indicated by the \textit{Error} label) as a fraction of the total number of fault/error injection experiments, distinguished by orchestration failure types. 
For most fault/error injections that lead to erroneous data store states and failures, the user does not receive any warning or error notification from the Apiserver. The ``scale" workload shows that a significant number of experiments return errors. The reason is that the workload generates multiple successive requests related to the same resource instances. In this context, a fault/error may compromise the resource integrity, triggering errors in subsequent requests.
In the case of the ``deploy" and ``failover" workloads, the majority of errors ($29/32$) were due to injections in the serialization protocol that prevented decoding and operations on the object.
The reason for the small number of errors is the delayed reconciliation of the cluster state and the desired state on the data store (see \S\ref{sec:intro}). The Apiserver only acknowledges receiving a request for a modification of values in Etcd. This does not imply that the cluster state changed.

\subsubsection{Injection propagation analysis}\label{subsec:propagation}\hfill

We performed a batch of experiments to understand which \revision{injected data values coming from other components} can be detected and blocked by the Apiserver validation layer. We used Mutiny to inject bit-flips into the messages sent \revision{towards the Apiserver} by the Kcm, the Scheduler, and the Kubelet, and \revision{we checked through logging} whether the injected values reached Etcd. The experiment was also intended to indicate the robustness of the client input validation. \revision{The experimental setup is the same of the campaign described in \S\ref{subsec:results}.}
\begin{table}[h]
\resizebox{\columnwidth}{!}{%
\begin{tabular}{l|lll|lll|lll|}
\cline{2-10}
 &
  \multicolumn{3}{c|}{Kcm} &
  \multicolumn{3}{c|}{Scheduler} &
  \multicolumn{3}{c|}{Kubelet} \\ \cline{2-10} 
 &
  \multicolumn{1}{l|}{Inj.} &
  \multicolumn{1}{l|}{Prop} &
  Err. &
  \multicolumn{1}{l|}{{\color[HTML]{2E3436} Inj.}} &
  \multicolumn{1}{l|}{{\color[HTML]{2E3436} Prop}} &
  {\color[HTML]{2E3436} Err.} &
  \multicolumn{1}{l|}{{\color[HTML]{2E3436} Inj.}} &
  \multicolumn{1}{l|}{{\color[HTML]{2E3436} Prop}} &
  {\color[HTML]{2E3436} Err.} \\ \hline
\multicolumn{1}{|l|}{Deploy} &
  \multicolumn{1}{l|}{468} &
  \multicolumn{1}{l|}{165} &
  38 &
  \multicolumn{1}{l|}{40} &
  \multicolumn{1}{l|}{9} &
  8 &
  \multicolumn{1}{l|}{69} &
  \multicolumn{1}{l|}{34} &
  7 \\ \hline
\multicolumn{1}{|l|}{Scale} &
  \multicolumn{1}{l|}{472} &
  \multicolumn{1}{l|}{165} &
  26 &
  \multicolumn{1}{l|}{40} &
  \multicolumn{1}{l|}{9} &
  8 &
  \multicolumn{1}{l|}{69} &
  \multicolumn{1}{l|}{31} &
  5 \\ \hline
\multicolumn{1}{|l|}{Failover} &
  \multicolumn{1}{l|}{382} &
  \multicolumn{1}{l|}{136} &
  29 &
  \multicolumn{1}{l|}{30} &
  \multicolumn{1}{l|}{9} &
  6 &
  \multicolumn{1}{l|}{69} &
  \multicolumn{1}{l|}{29} &
  5 \\ \hline
\end{tabular}%
}
\caption{Results of the propagation experiments. \textit{Inj.}: the number of injections performed. \textit{Prop}: values propagated to Etcd.  \revision{\textit{Err.}: an error was logged due to  a wrong value.} The Kcm has more injections performed because it manipulates multiple types of resources and updates larger numbers of fields than the other two components.}
\label{tab:propagation}
\end{table}

The fault/error injections targeted the same set of message fields as our previous injection campaign.
\tableautorefname~\ref{tab:propagation} shows the results from those experiments.
We observed that for $12$ of the critical data fields (previously analyzed), the validation mechanisms failed to capture data corruption, i.e., the corrupted inputs propagated from the Kcm without being intercepted and caused failures like SU. The reason is that the Apiserver performs general validations, e.g., regex matching or border-case testing, but is not able to detect valid but wrong values. 
Fortunately, the Apiserver is able to validate some of the fields that can cause severe error patterns. For example, the Apiserver detects and prevents the propagation of a namespace that does not match the URL or label selectors that do not match the template labels of the same resource instance, a condition that triggers the infinite Pod spawn.
Values from the Kubelet, which may be more likely to be compromised, impact only the Node itself or a single Pod, presenting a low risk.

%% file: 6_discussion.tex
\subsection{How can we improve testing?}
\begin{table}[]
\begin{minipage}[t]{\columnwidth}
\begin{adjustbox}{height=120pt}
\begin{tabular}{|p{0.17\textwidth}|p{1.16\textwidth}|p{0.01\textwidth}}
\hline
\rowcolor{Gray}
\textbf{Error} & \textbf{Error Subcategories} \\
\hline
State Retrieval & State corrupted, erased, stale, unretrievable \\
\hline
Misbehav. Logic & \textbf{Wrong label}, \textbf{Wrong replica value}, \textbf{Request rejected}, \textbf{Lost update}, \textbf{Controller loop not executed}, \textbf{Relationship broken}\\
\hline
Commun. Problems & Connection delay, \textbf{Wrong IP address}, DNS resolution delay, \textbf{DNS not resolving}, Uneven load balancing, Endpoint delete after Pod kill,  \textbf{Routes dropped}, \textbf{\textit{New Nodes' routes not configured}}, \textbf{\textit{Routes not updated}}\\
\hline
Capacity \newline Exceed & \textbf{Overcrowding}, \textbf{Cluster out of resources}, \textbf{Worker nodes cannot join}, \textbf{Worker nodes unhealthy} \\
\hline
CP Availab. & \textbf{CP Pods crash loop}, \textit{\textbf{CP Pods hang}}, CP Pods deleted, \textbf{CP overload}\\
\hline
Local to Nodes & Kubelet delayed, Container runtime failure, \textbf{Pods not ready}, \textbf{Image Pull Error}, Slow/throttling\\
\hline
\hline
\rowcolor{Gray}
\textbf{Failure} & \textbf{Failure Subcategories} \\
\hline
Cluster \newline Outage & \textbf{Cluster-wide networking drop}, Cluster-wide networking intermittent, 
\textbf{Massive Service Deletion},
\textbf{DNS resolution failure}\\
\hline
Stall & \textbf{Control Plane stuck}, \textbf{Control Plane slow}, Control Plane quorum unreachable, \textbf{New Services network not configurable}, \textit{\textbf{New Nodes network not reconfigurable}}\\
\hline
Service \newline Networking & \textbf{Service Networking Drop Permanent}, \textbf{Service Networking Drop Intermittent}, Service Networking Delay\\
\hline
More Res. & \textbf{Pods not deleted}, \textbf{Too many Pods created},  \textbf{\textit{More Pods Transient}}, \textbf{More Resources Per Pod}\\
\hline
Less Res. & \textbf{Pods deleted}, \textbf{Pods not created}, \textbf{Pods crashloop}, \textbf{Less Resources Per Pod}\\
\hline
Timing & \textbf{Pods' Creation Delayed}, \textit{\textbf{Pods Restart}}\\
\hline
\end{tabular}%
\end{adjustbox}
\end{minipage}
\caption{Comparison between injections and the real world. \textbf{Bold} indicates what Mutiny can replicate, and \textit{italics} indicates what is triggered by Mutiny and not present in the real world.}
\label{tab:replication}
\vspace{-10pt}
\end{table}

\tableautorefname~\ref{tab:replication} shows failures triggered by Mutiny compared to failures reported in the real world. Our injector easily triggers errors related to logic, capacity, and control plane availability. On the other hand, it fails to trigger several errors local to the worker Nodes, because those errors are mainly due to local configurations and underlying software (e.g., kernel, runtimes) problems. 
For example, it falls short in inducing delays caused by DNS resolution, connection errors, arbitrary numbers (different from 1 and \textit{all}) of unhealthy Nodes, and transient and intermittent network failures in general. 
Nonetheless, almost all failure subcategories can be covered. 
At any rate, our aim is not to create a one-size-fits-all injector, but rather to provide a framework capable of triggering unforeseen error patterns to test the system response and provide insights that can be used to devise methods for mitigating or recovering from potential failures. 
Currently, companies use well-known techniques for testing applications' resiliency (e.g., Chaos engineering), but there is limited understanding of how to test the resiliency of the platform itself. At the moment, the responsibility for setting up proper monitoring alerts to detect failures and enforce manual mitigation belongs to cluster operators. Sometimes custom code is employed to prevent past failures from happening again; for example, validation of namespace deletion can prevent accidental deletion of a non-empty namespace together with all its Pods.
However, a post-incident manual procedure or code customization cannot be the answer to a trend that sees the use of K8s as a cloud OS for more critical scenarios with tight non-functional requirements. 

Systematic orchestration resiliency tests should become an integral part of the development process to get quantitative metrics of resiliency strategies in place. 
A set of injectors, such as network and OS-level injectors, can be used in testing clusters in a Chaos-engineering fashion to train human operators and improve handbooks containing procedures to follow when a failure occurs.
With a real workload, Mutiny would introduce errors in a set of events that rarely happen, but that can be troublesome (e.g., updates of networking \textit{ConfigMaps}, certificates, and secrets). This would trigger new failure patterns not reflected in our experiments, due to simple workloads. 
For each critical failure pattern, appropriate and systematic countermeasures should be designed before deployment in production environments.

\revision{Findings like F1 and the systematic orchestration resiliency tests are not specific to K8s: they descend from the inherent problem that these systems control resources at a huge scale. } \cameraready{The proposed fault-injection methodology can be ported to other orchestration systems, as they all share common architectural principles: the study in \cite{truyen2019comprehensive} compares several orchestration frameworks (e.g., K8S, Docker Swarm, Mesos, Autora, Marathon), observing that all of them keep applications in their desired state by comparing the monitored state and the desired state. Moreover, controllers use a data store (e.g., Etcd, Consul, or Zookeeper) for storing the states.}



\subsection{What can we do about failures?}
$\blacksquare$ \textbf{ The system design} can be improved in terms of data validation for critical resources and fields. In K8s there is a massive use of labels because of their flexibility in grouping and selecting resources. We showed that this flexibility comes at the expense of resiliency, because it is hard to validate their custom values. Recall the Reddit \cite{redditpiday} failure discussed earlier, in which a single label tore down the entire cluster network. Hence, updates to critical fields and resources (e.g., control plane Pods or Nodes) should be logged. \revision{In case of logging subsystem failure, additional data sources (e.g., Prometheus) can be used to detect the change}. Upon a change, system behavior should be monitored to detect any degradation of the system's health, so it is possible to roll back changes to critical fields. Moreover, stricter checks can be enforced: e.g., scaling of coreDNS to 0 should be denied, while adding the \textit{MaxUnavailability} parameter could prevent outages.
Logs can be used to derive a nominal behavior of the system. \revision{Mutiny can be used to conduct a log analysis to check what K8s logs under injection, to possibly improve the logging when no traces about failures/errors are found. To this aim, K8s auditing (currently a beta feature \cite{securing}) can be used together with injections. Then,} 
methods like model checking and runtime verification, \revision{can} be used to validate data exchanged between components and detect anomalies. 
\cameraready{Log analysis can be used to derive dependability measurements for each K8s component and assess failure propagation patterns among them. The analysis can help to spot the most problematic components. For instance, our experiments showed that components running validations or controlling scaling and cluster-wide configurations (i.e., Apiserver, Kcm, and network managers) can be troublesome, other than the centralized state itself which represents the main dependability bottleneck.}

Although the analysis in \S\ref{subsec:propagation} reveals that data-validation mechanisms can prevent some severe failures, it is not enough to validate the data only once. If for some reason an incorrect value gets to Etcd (perhaps introduced by the Apiserver itself), escaping data validation, no circuit breaker, or other resiliency strategies mitigate the impact on the system. Real-world incidents have proven that incorrect data can escape data validation, causing, for example, uncontrolled replication of resources. 
Circuit breakers must thus be systematically designed to cover all the resource kinds that can cause overload errors, for example, when the relationship between resource instances is broken.
\revision{Furthermore, when writing custom controllers, the replication management must be designed resilient to a variety of faults/errors.}
\revision{Simple data redundancy mechanisms, like redundancy codes on critical fields, can protect the cluster from hardware faults with a negligible overhead in terms of resource usage (the critical fields are  $<10\%$ of total).}

$\blacksquare$ \textbf{Cluster managers},
\revision{ since a single error can disrupt an entire cluster, should prefer multiple clusters rather than a reduced number of high-scale clusters to contain error propagation; at the cost of additional control plane Nodes and management complexity.
Cluster managers should set and test upper resource limits in terms of Pod resources, number of Nodes in the cluster, request rates, and number of spawned Pods.
\cameraready{\textit{Namespace} features can limit resource counts and quotas\cite{resourcequotas} to somehow partition different tenant/service types and mitigate failures.}
Cluster managers must be aware of the default parameters and the software specification in non-nominal conditions. 
Mechanisms like \textit{MaxUnavailability}, \textit{MaxSurge} (i.e., maximum Pod number that can be created over the desired one), and backoff timers should be tuned thinking at failures, despite slowing down daily operations. }
\revision{From a security perspective, the access to Etcd must be strictly guarded by authentication and firewalls. K8s configures Etcd with client authentication, but not rarely administrators directly connect to Etcd. K8s security features can reduce the attack surface from unauthorized users, but cannot prevent errors generated in the authorized clients, e.g., Apiserver or Kcm.}

Capacity-related failures in real-world incidents have been caused by misconfigurations and human mistakes more often than by internal system errors. User requests that can overload the system should be blocked, e.g., reject the spawning of a large number of Pods without resource limits 
or slow down evictions due to preemption by a new deployment.

%% file: 7_related.tex
\subsection{Dependability assessment for cloud systems}
Software and hardware injections have been debated for a long time \cite{hsueh1997fault, natella2016assessing, moraes2006injection}.
The authors of \cite{tang2023fail} analyze production failures in cloud systems, arguing that they cannot be understood by analyzing a single system in isolation. Hence, they introduce the concept of cross-system interaction failures, and discuss potential mitigations. 
The authors of \cite{di2014lessons} use FFDA to analyze the failures of the Blue Waters supercomputer.  
The authors of \cite{viazarreta2020dependability} describe dependability bottlenecks through stochastic models of two software-defined network (SDN) controllers. SDNs feature a distributed control plane, like container orchestration systems. Several papers have studied the availability assessment for cloud and edge computing \cite{maciel2021survey,khazaei2012availability,ghosh2014scalable,de2022latency,faraji2021availability}. The most common approaches use analytical models like Markov chains or Petri nets, and rely on field measurements.

\subsection{Fault injection in cloud systems}
The authors of \cite{sun2022automatic} created a testing framework for custom K8s controllers by perturbing the controller’s view of the current cluster state through stale-state, crash, and unobserved states.
That was followed by \cite{gu2023acto}, which describes how end-to-end tests can be generated to trigger state changes from states different from the initial one.
Oracle state checking is used to detect misbehavior. 
The approach described in \cite{lu2019crashtuner} looks for crash-recovery bugs in distributed systems by injecting crashes at precise points identified through the analysis of meta-information used by nodes. Similarly, in \cite{liu2018fcatch}, time-of-fault bugs 
are found by identifying conflicting operations based on correct runs, and exploiting the ones not covered by fault-tolerance mechanisms.
The authors of \cite{chen2020cofi} injected network partitions to discover partition bugs. The injections are performed when consistency invariants are violated. 
Simple random partitions were found to be useful in \cite{majumdar2018random} in which the probability bounds for discovering bugs are derived.
The authors of \cite{sun2021reasoning} argue that partial histories, including staleness, time traveling, and observability gaps, are an inherent threat to distributed systems that locally cache their state, as K8s does. 

The above work all aimed to expose flaws by leveraging the inherent weaknesses of distributed systems and injecting simple faults (e.g., crash and stale states) in strategic ways. 
Unlike previous studies of operating systems \cite{gu2003characterization,jarboui2002experimental,cotroneo2009assessment} and past cloud platforms \cite{cotroneo2019bad,ju2013fault}, the above publications do not provide a systematic assessment of failures and resiliency. 
Unlike them, we focus on fault tolerance, assuming the presence of residual bugs or even components that are flawless but can nevertheless lead to system failures.
The authors of  \cite{flora2022study} used fault injection 
to study the effectiveness of K8s at handling the aging and faults of deployed microservices, and concluded that probes fall short in detecting several failure modes. 
In 2016, Netflix introduced Chaos engineering \cite{basiri2016chaos}, which automatically, randomly, and deliberately introduces faults through injections in production systems to find and improve dependability bottlenecks. \revision{Chaos engineering effectiveness relies on the simple fault/error model that can be applied without being aware of services' semantics, highlighting the bottlenecks of a complex system topology.}
Unlike us, those efforts focused on the resiliency of deployed services (\S\ref{subsec:fault-tolerance}), while we claim that chaos-engineering--like methods should be applied to the \revision{components of the} orchestration system itself, \revision{which are fewer and with known interaction patterns. This allows taking advantage of the architecture to inject tailored faults/errors.}
\subsection{Fault-tolerant designs in orchestration systems}
Papers \cite{shahid2021towards,kumari2021survey} provide reviews of fault-tolerance methods in cloud environments.
In \cite{diouf2020byzantine}, Byzantine fault tolerance is integrated into the K8s control plane through state machine replication. The work in \cite{sakic2019p4bft} does something similar for SDN control planes, while \cite{netto2017state} introduces state machine replications for the applications deployed in K8s. However, state machine replication cannot mitigate common cause failures, e.g., deterministic failures due to misconfigurations, mistakes, bugs, and upgrades.
The authors of \cite{zhou2021tardis} describe how violations of invariants can be used to detect deterministic bugs, and how semantic-equivalent input transformation through symbolic execution can be performed to recover dynamically.

%% file: 7_conclusion.tex
\section{Conclusion}
\label{sec:conclusion}
We classified real-world incidents to analyze how Kubernetes fails, and we described a fault injection campaign we performed that altered the data representing the cluster state to reproduce some incidents and also trigger new error patterns. We introduced a failure model for orchestration systems that we used to analyze our experimental results. Although K8s resiliency strategies can tolerate quite a lot of errors, the system is sensitive to state alterations, and a single bad value can cause overloads and cluster-wide failures. The mechanisms enabling flexible relationships among resource instances can cause such critical failures. Hence, injection-based testing is essential to proactively assess orchestration resiliency, guide the design of failure mitigation actions, and set up monitoring alerts. Nonetheless, a systematic design of resiliency mechanisms is required to prevent system-wide failures. 
\section*{Acknowledgments}
We thank the reviewers, S. Cui, H. Qiu,  H. Sreejith, A. Patke, P. Cao, J. Applequist, and K. Atchley for the insightful comments on the early drafts.
We acknowledge the early participation of Larisa Shwartz (IBM) and Saurabh Jha (IBM) in the conceptualization of fault injection methods for Kubernetes; and Chandra Narayanaswami (IBM) for his continued insights and support on related system issues.
This work is partially supported by the National Science Foundation (NSF) under grant No. 2029049; by the  IBM-ILLINOIS Discovery Accelerator Institute (IIDAI); a gift from Nokia Bell Labs Core Research; and by the Italian Ministry of Enterprises and Made in Italy (MIMIT) under the GENIO Project (CUP B69J23005770005). In memory of Fabio Barletta.

%% file: main.bbl

%% file: main.bbl
\begin{thebibliography}{10}
\providecommand{\url}[1]{#1}
\csname url@samestyle\endcsname
\providecommand{\newblock}{\relax}
\providecommand{\bibinfo}[2]{#2}
\providecommand{\BIBentrySTDinterwordspacing}{\spaceskip=0pt\relax}
\providecommand{\BIBentryALTinterwordstretchfactor}{4}
\providecommand{\BIBentryALTinterwordspacing}{\spaceskip=\fontdimen2\font plus
\BIBentryALTinterwordstretchfactor\fontdimen3\font minus \fontdimen4\font\relax}
\providecommand{\BIBforeignlanguage}[2]{{%
\expandafter\ifx\csname l@#1\endcsname\relax
\typeout{** WARNING: IEEEtran.bst: No hyphenation pattern has been}%
\typeout{** loaded for the language `#1'. Using the pattern for}%
\typeout{** the default language instead.}%
\else
\language=\csname l@#1\endcsname
\fi
#2}}
\providecommand{\BIBdecl}{\relax}
\BIBdecl

\bibitem{khan2017key}
A.~Khan, ``Key characteristics of a container orchestration platform to enable a modern application,'' \emph{IEEE cloud Computing}, vol.~4, no.~5, pp. 42--48, 2017.

\bibitem{rodriguez2019orchestration}
M.~A. Rodriguez and R.~Buyya, ``Container-based cluster orchestration systems: A taxonomy and future directions,'' \emph{Software: Practice and Experience}, vol.~49, no.~5, pp. 698--719, 2019.

\bibitem{burns2016borg}
B.~Burns, B.~Grant, D.~Oppenheimer, E.~Brewer, and J.~Wilkes, ``Borg, omega, and kubernetes: Lessons learned from three container-management systems over a decade,'' \emph{Queue}, vol.~14, no.~1, pp. 70--93, 2016.

\bibitem{vxworks}
{Windriver}. {Containers at the Intelligent Edge}. \url{https://www.windriver.com/resource/containers-at-the-intelligent-edge}. Accessed \today.

\bibitem{mellado2022design}
J.~Mellado and F.~N{\'u}{\~n}ez, ``{Design of an IoT-PLC: A containerized programmable logical controller for the industry 4.0},'' \emph{Journal of Industrial Information Integration}, vol.~25, p. 100250, 2022.

\bibitem{barletta2023criticality}
M.~Barletta, M.~Cinque, L.~De~Simone, and R.~D. Corte, ``Criticality-aware monitoring and orchestration for containerized industry 4.0 environments,'' \emph{ACM Transactions on Embedded Computing Systems}, vol.~23, no.~1, pp. 1--28, 2023.

\bibitem{johansson22kubernetes}
B.~Johansson, M.~R{\aa}gberger, T.~Nolte, and A.~V. Papadopoulos, ``Kubernetes orchestration of high availability distributed control systems,'' in \emph{2022 IEEE International Conference on Industrial Technology (ICIT)}.\hskip 1em plus 0.5em minus 0.4em\relax IEEE, 2022, pp. 1--8.

\bibitem{nehpio}
{Linux Foundation}. {Nephio: Cloud Native Network Automation}. \url{https://nephio.org/about/}. Accessed \today.

\bibitem{ferguson2023corekube}
A.~E. Ferguson, J.~Larrea, and M.~K. Marina, ``Corekube: An efficient, autoscaling and resilient mobile core system,'' in \emph{The 29th Annual International Conference On Mobile Computing And Networking}.\hskip 1em plus 0.5em minus 0.4em\relax ACM Association for Computing Machinery, 2023, pp. 1--15.

\bibitem{botez2021sdn}
R.~Botez, J.~Costa-Requena, I.-A. Ivanciu, V.~Strautiu, and V.~Dobrota, ``{SDN-based network slicing mechanism for a scalable 4G/5G core network: A kubernetes approach},'' \emph{Sensors}, vol.~21, no.~11, p. 3773, 2021.

\bibitem{colombo2010factory}
A.-W. Colombo, S.~Karnouskos, and J.-M. Mendes, ``Factory of the future: A service-oriented system of modular, dynamic reconfigurable and collaborative systems,'' in \emph{Artificial intelligence techniques for networked manufacturing enterprises management}.\hskip 1em plus 0.5em minus 0.4em\relax Springer, 2010, pp. 459--481.

\bibitem{morgan2021industry}
J.~Morgan, M.~Halton, Y.~Qiao, and J.~G. Breslin, ``Industry 4.0 smart reconfigurable manufacturing machines,'' \emph{Journal of Manufacturing Systems}, vol.~59, pp. 481--506, 2021.

\bibitem{lasi2014industry}
H.~Lasi, P.~Fettke, H.-G. Kemper, T.~Feld, and M.~Hoffmann, ``Industry 4.0,'' \emph{Business \& information systems engineering}, vol.~6, pp. 239--242, 2014.

\bibitem{ray2019edge}
P.~P. Ray, D.~Dash, and D.~De, ``Edge computing for internet of things: A survey, e-healthcare case study and future direction,'' \emph{Journal of Network and Computer Applications}, vol. 140, pp. 1--22, 2019.

\bibitem{certifiedkube}
{Cloud Native Computing Foundation}. (2022) {Kubernetes Certified Distributions}. \url{https://www.cncf.io/certification/software- conformance/}. Accessed \today.

\bibitem{cncf-survey}
C.~N.~C. Foundation, ``{CNCF Annual Survey 2021},'' \url{https://www.cncf.io/reports/cncf-annual-survey-2021/}, 2022, accessed \today.

\bibitem{kubecon5}
L.~Bernaille and R.~Boll, ``{10 Ways to Shoot Yourself in the Foot with Kubernetes},'' \url{https://www.youtube.com/watch?v=QKI-JRs2RIE}, 2020, accessed \today.

\bibitem{k8sfailures}
hjacobs, ``{Kubernetes Failure Stories},'' \url{https://k8s.af/}, 2023, accessed \today.

\bibitem{kubecon1}
Airbnb, ``{10 More Weird Ways to Blow Up Your Kubernetes},'' \url{https://www.youtube.com/watch?v=4CT0cI62YHk}, 2021, accessed \today.

\bibitem{kubecon2}
S.~Visvanathan and N.~Venkatachalam, ``{101 Ways to “Break and Recover” Kubernetes Cluster},'' \url{https://www.youtube.com/watch?v=likHm-KHGWQ}, 2018, accessed \today.

\bibitem{redditpiday}
J.~Howard, ``{You Broke Reddit: The Pi-Day Outage},'' \url{https://www.reddit.com/r/RedditEng/comments/11xx5o0/you_broke_reddit_the_piday_outage/}, 2023, accessed \today.

\bibitem{failuregke}
Venafi, ``{How a Simple Kubernetes Admission Webhook Lead to a Cluster Outage},'' \url{https://venafi.com/blog/gke-webhook-outage/}, 2019, accessed \today.

\bibitem{gkefailures}
\BIBentryALTinterwordspacing
{Google}. (2022) {All incidents reported for Google Kubernetes Engine }. [Online]. Available: \url{https://status.cloud.google.com/products/LCSbT57h59oR4W98NHuz/history}
\BIBentrySTDinterwordspacing

\bibitem{repo}
M.~Barletta, ``{Mutiny},'' \url{https://dessert.unina.it:8088/marcobarlo/mutiny} "Mutiny-Scripts" \url{https://dessert.unina.it:8088/marcobarlo/mutiny-scripts}, note = "Accessed \today",, 2023.

\bibitem{sun2020testing}
X.~Sun, R.~Cheng, J.~Chen, E.~Ang, O.~Legunsen, and T.~Xu, ``Testing configuration changes in context to prevent production failures,'' in \emph{14th USENIX Symposium on Operating Systems Design and Implementation (OSDI 20)}, 2020, pp. 735--751.

\bibitem{barletta2022sla}
M.~Barletta, M.~Cinque, and C.~Di~Martino, ``{SLA-Driven Software Orchestration in Industry 4.0},'' \emph{IEEE Internet of Things Magazine}, vol.~5, no.~4, pp. 136--141, 2022.

\bibitem{sun2022automatic}
X.~Sun, W.~Luo, J.~T. Gu, A.~Ganesan, R.~Alagappan, M.~Gasch, L.~Suresh, and T.~Xu, ``Automatic reliability testing for cluster management controllers,'' in \emph{16th USENIX Symposium on Operating Systems Design and Implementation (OSDI 22)}, 2022, pp. 143--159.

\bibitem{gu2023acto}
J.~T. Gu, X.~Sun, W.~Zhang, Y.~Jiang, C.~Wang, M.~Vaziri, O.~Legunsen, and T.~Xu, ``Acto: Automatic end-to-end testing for operation correctness of cloud system management,'' in \emph{Proceedings of the 29th Symposium on Operating Systems Principles}, 2023, pp. 96--112.

\bibitem{lu2019crashtuner}
J.~Lu, C.~Liu, L.~Li, X.~Feng, F.~Tan, J.~Yang, and L.~You, ``Crashtuner: detecting crash-recovery bugs in cloud systems via meta-info analysis,'' in \emph{Proceedings of the 27th ACM Symposium on Operating Systems Principles (SOSP)}, 2019, pp. 114--130.

\bibitem{liu2018fcatch}
H.~Liu, X.~Wang, G.~Li, S.~Lu, F.~Ye, and C.~Tian, ``{FCatch: Automatically detecting time-of-fault bugs in cloud systems},'' \emph{ACM SIGPLAN Notices}, vol.~53, no.~2, pp. 419--431, 2018.

\bibitem{chen2020cofi}
H.~Chen, W.~Dou, D.~Wang, and F.~Qin, ``Cofi: consistency-guided fault injection for cloud systems,'' in \emph{Proceedings of the 35th IEEE/ACM International Conference on Automated Software Engineering}, 2020, pp. 536--547.

\bibitem{sun2021reasoning}
X.~Sun, L.~Suresh, A.~Ganesan, R.~Alagappan, M.~Gasch, L.~Tang, and T.~Xu, ``Reasoning about modern datacenter infrastructures using partial histories,'' in \emph{Proceedings of the Workshop on Hot Topics in Operating Systems}, 2021, pp. 213--220.

\bibitem{majumdar2018random}
R.~Majumdar and F.~Niksic, ``Why is random testing effective for partition tolerance bugs?'' \emph{Proceedings of the ACM on Programming Languages}, vol.~2, pp. 1--24, 2017.

\bibitem{meng2023distributed}
R.~Meng, G.~P{\^\i}rlea, A.~Roychoudhury, and I.~Sergey, ``Distributed system fuzzing,'' \emph{arXiv preprint arXiv:2305.02601}, 2023.

\bibitem{basiri2016chaos}
A.~Basiri, N.~Behnam, R.~De~Rooij, L.~Hochstein, L.~Kosewski, J.~Reynolds, and C.~Rosenthal, ``Chaos engineering,'' \emph{IEEE Software}, vol.~33, no.~3, pp. 35--41, 2016.

\bibitem{flora2022study}
J.~Flora, P.~Gon{\c{c}}alves, M.~Teixeira, and N.~Antunes, ``A study on the aging and fault tolerance of microservices in kubernetes,'' \emph{IEEE Access}, vol.~10, pp. 132\,786--132\,799, 2022.

\bibitem{saurabh400firm}
H.~Qiu, S.~S. Banerjee, S.~Jha, Z.~T. Kalbarczyk, and R.~K. Iyer, ``{FIRM}: An intelligent fine-grained resource management framework for slo-oriented microservices,'' in \emph{14th {USENIX} Symposium on Operating Systems Design and Implementation ({OSDI} 20)}.\hskip 1em plus 0.5em minus 0.4em\relax Banff, Alberta: {USENIX} Association, Nov. 2020, pp. 805--825.

\bibitem{kubecon7}
H.~Jacobs, ``{Kubernetes Failure Stories, or: How to Crash Your Cluster},'' \url{https://www.youtube.com/watch?v=LpFApeaGv7A}, 2019.

\bibitem{ongaro2014search}
D.~Ongaro and J.~Ousterhout, ``In search of an understandable consensus algorithm,'' in \emph{2014 USENIX annual technical conference (USENIX ATC 14)}, 2014, pp. 305--319.

\bibitem{leveltriggered}
J.~Bowes, ``{Level Triggering and Reconciliation in Kubernetes},'' \url{https://hackernoon.com/level-triggering-and-reconciliation-in-kubernetes-1f17fe30333d}, 2020, accessed \today.

\bibitem{failureKubelet}
Zalando, ``{Let's talk about Failures with Kubernetes - Hamburg Meetup},'' \url{https://www.slideshare.net/try_except_/lets-talk-about-failures-with-kubernetes-hamburg-meetup}, 2019, accessed \today.

\bibitem{issuekbue}
K.~repository users, ``{Kubernetes Issue 69579 },'' \url{https://github.com/kubernetes/kubernetes/issues/69579}, 2018, accessed \today.

\bibitem{kubecon3}
M.~Cebula and B.~S. Airbnb, ``{10 Weird Ways to Blow Up Your Kubernetes},'' \url{https://www.youtube.com/watch?v=FrQ8Lwm9_j8}, 2020, accessed \today.

\bibitem{dolev1996failure}
D.~Dolev, R.~Friedman, I.~Keidar, and D.~Malkhi, ``Failure detectors in omission failure environments,'' Cornell University, Tech. Rep., 1996.

\bibitem{DeathStartBench}
delimitrou, ``{DeathStarBench},'' \url{https://github.com/delimitrou/DeathStarBench}, accessed \today.

\bibitem{kbench}
{VMware Tanzu}. {k-bench}. \url{https://github.com/vmware-tanzu/k-bench}. Accessed \today.

\bibitem{components}
Kubernetes, ``Kubernetes docs - components,'' \url{https://kubernetes.io/docs/concepts/overview/components/}, 2023.

\bibitem{truyen2019comprehensive}
E.~Truyen, D.~Van~Landuyt, D.~Preuveneers, B.~Lagaisse, and W.~Joosen, ``A comprehensive feature comparison study of open-source container orchestration frameworks,'' \emph{Applied Sciences}, 2019.

\bibitem{securing}
Kubernetes, ``Kubernetes docs - scuring a cluster,'' \url{https://kubernetes.io/docs/tasks/administer-cluster/securing-a-cluster/}, 2023.

\bibitem{resourcequotas}
K.~developers, ``{Resource quotas},'' \url{https://kubernetes.io/docs/concepts/policy/resource-quotas/}, 2024, accessed \today.

\bibitem{hsueh1997fault}
M.-C. Hsueh, T.~K. Tsai, and R.~K. Iyer, ``Fault injection techniques and tools,'' \emph{Computer}, vol.~30, no.~4, pp. 75--82, 1997.

\bibitem{natella2016assessing}
R.~Natella, D.~Cotroneo, and H.~S. Madeira, ``Assessing dependability with software fault injection: A survey,'' \emph{ACM Computing Surveys (CSUR)}, vol.~48, no.~3, pp. 1--55, 2016.

\bibitem{moraes2006injection}
R.~Moraes, R.~Barbosa, J.~Dur{\~a}es, N.~Mendes, E.~Martins, and H.~Madeira, ``Injection of faults at component interfaces and inside the component code: are they equivalent?'' in \emph{Proceedings of Sixth European Dependable Computing Conference (EDCC)}.\hskip 1em plus 0.5em minus 0.4em\relax IEEE, 2006, pp. 53--64.

\bibitem{tang2023fail}
L.~Tang, C.~Bhandari, Y.~Zhang, A.~Karanika, S.~Ji, I.~Gupta, and T.~Xu, ``Fail through the cracks: Cross-system interaction failures in modern cloud systems,'' in \emph{Proceedings of the Eighteenth European Conference on Computer Systems}, 2023, pp. 433--451.

\bibitem{di2014lessons}
C.~Di~Martino, Z.~Kalbarczyk, R.~K. Iyer, F.~Baccanico, J.~Fullop, and W.~Kramer, ``Lessons learned from the analysis of system failures at petascale: The case of blue waters,'' in \emph{2014 44th Annual IEEE/IFIP International Conference on Dependable Systems and Networks}.\hskip 1em plus 0.5em minus 0.4em\relax IEEE, 2014, pp. 610--621.

\bibitem{viazarreta2020dependability}
P.~Vizarreta, K.~Trivedi, V.~Mendiratta, W.~Kellerer, and C.~Mas-Machuca, ``Dason: Dependability assessment framework for imperfect distributed sdn implementations,'' \emph{IEEE Transactions on Network and Service Management}, vol.~17, no.~2, pp. 652--667, 2020.

\bibitem{maciel2021survey}
P.~Maciel, J.~Dantas, C.~Melo, P.~Pereira, F.~Oliveira, J.~Araujo, and R.~Matos, ``A survey on reliability and availability modeling of edge, fog, and cloud computing,'' \emph{Journal of Reliable Intelligent Environments}, pp. 1--19, 2021.

\bibitem{khazaei2012availability}
H.~Khazaei, J.~Mi{\v{s}}i{\'c}, V.~B. Mi{\v{s}}i{\'c}, and N.~B. Mohammadi, ``Availability analysis of cloud computing centers,'' in \emph{Proceedings of IEEE Global Communications Conference (GLOBECOM)}.\hskip 1em plus 0.5em minus 0.4em\relax IEEE, 2012, pp. 1957--1962.

\bibitem{ghosh2014scalable}
R.~Ghosh, F.~Longo, F.~Frattini, S.~Russo, and K.~S. Trivedi, ``{Scalable analytics for IaaS cloud availability},'' \emph{IEEE Transactions on Cloud Computing}, vol.~2, no.~1, pp. 57--70, 2014.

\bibitem{de2022latency}
L.~De~Simone, M.~Di~Mauro, R.~Natella, and F.~Postiglione, ``A latency-driven availability assessment for multi-tenant service chains,'' \emph{IEEE Transactions on Services Computing}, vol.~16, no.~2, pp. 815--829, 2022.

\bibitem{faraji2021availability}
M.~Faraji~Shoyari, E.~Ataie, R.~Entezari-Maleki, and A.~Movaghar, ``Availability modeling in redundant openstack private clouds,'' \emph{Software: Practice and Experience}, vol.~51, no.~6, pp. 1218--1241, 2021.

\bibitem{gu2003characterization}
W.~Gu, Z.~Kalbarczyk, R.~K. Iyer, and Z.~Yang, ``Characterization of linux kernel behavior under errors,'' in \emph{Proceedings of International Conference on Dependable Systems and Networks}.\hskip 1em plus 0.5em minus 0.4em\relax IEEE Computer Society, 2003, pp. 459--459.

\bibitem{jarboui2002experimental}
T.~Jarboui, J.~Arlat, Y.~Crouzet, and K.~Kanoun, ``Experimental analysis of the errors induced into linux by three fault injection techniques,'' in \emph{Proceedings of the 2002 IEEE International Conference on Dependable Systems and Networks (DSN)}.\hskip 1em plus 0.5em minus 0.4em\relax IEEE, 2002, pp. 331--336.

\bibitem{cotroneo2009assessment}
D.~Cotroneo, R.~Natella, and S.~Russo, ``Assessment and improvement of hang detection in the linux operating system,'' in \emph{2009 28th IEEE International Symposium on Reliable Distributed Systems}.\hskip 1em plus 0.5em minus 0.4em\relax IEEE, 2009, pp. 288--294.

\bibitem{cotroneo2019bad}
D.~Cotroneo, L.~De~Simone, P.~Liguori, R.~Natella, and N.~Bidokhti, ``How bad can a bug get? an empirical analysis of software failures in the openstack cloud computing platform,'' in \emph{Proceedings of the 2019 27th ACM Joint Meeting on European Software Engineering Conference and Symposium on the Foundations of Software Engineering}, 2019, pp. 200--211.

\bibitem{ju2013fault}
X.~Ju, L.~Soares, K.~G. Shin, K.~D. Ryu, and D.~Da~Silva, ``On fault resilience of openstack,'' in \emph{Proceedings of the 4th annual Symposium on Cloud Computing}, 2013, pp. 1--16.

\bibitem{shahid2021towards}
M.~A. Shahid, N.~Islam, M.~M. Alam, M.~Mazliham, and S.~Musa, ``Towards resilient method: An exhaustive survey of fault tolerance methods in the cloud computing environment,'' \emph{Computer Science Review}, vol.~40, p. 100398, 2021.

\bibitem{kumari2021survey}
P.~Kumari and P.~Kaur, ``A survey of fault tolerance in cloud computing,'' \emph{Journal of King Saud University-Computer and Information Sciences}, vol.~33, no.~10, pp. 1159--1176, 2021.

\bibitem{diouf2020byzantine}
G.~M. Diouf, H.~Elbiaze, and W.~Jaafar, ``On byzantine fault tolerance in multi-master kubernetes clusters,'' \emph{Future Generation Computer Systems}, vol. 109, pp. 407--419, 2020.

\bibitem{sakic2019p4bft}
E.~Sakic, N.~Deric, E.~Goshi, and W.~Kellerer, ``P4bft: Hardware-accelerated byzantine-resilient network control plane,'' in \emph{2019 IEEE Global Communications Conference (GLOBECOM)}.\hskip 1em plus 0.5em minus 0.4em\relax IEEE, 2019, pp. 1--7.

\bibitem{netto2017state}
H.~V. Netto, L.~C. Lung, M.~Correia, A.~F. Luiz, and L.~M.~S. de~Souza, ``State machine replication in containers managed by kubernetes,'' \emph{Journal of Systems Architecture}, 2017.

\bibitem{zhou2021tardis}
Z.~Zhou, T.~A. Benson, M.~Canini, and B.~Chandrasekaran, ``Tardis: A fault-tolerant design for network control planes,'' in \emph{Proceedings of the ACM SIGCOMM Symposium on SDN Research (SOSR)}, 2021, pp. 108--121.

\end{thebibliography}
